\documentclass[vecphys]{svmult}

\usepackage{makeidx}         
\usepackage{graphicx}        
\usepackage{multicol}        

\makeindex             

\begin{document}

\title*{Path Integrals in the Physics of Lattice Polarons}
\author{Pavel Kornilovitch\inst{}}
\institute{Hewlett-Packard, Corvallis, Oregon, 97330, USA \\
\texttt{pavel.kornilovich@hp.com}}
%
%
\maketitle

\begin{abstract}

A path-integral approach to lattice polarons is developed. The method is based on exact
analytical elimination of phonons and subsequent Monte Carlo simulation of self-interacting
fermions. The analytical basis of the method is presented with emphasis on visualization of 
polaron effects, which path integrals provide. Numerical results on the polaron energy, mass, 
spectrum and density of states are given for short-range and long-range electron-phonon 
interactions. It is shown that certain long-range interactions significantly reduce the 
polaron mass, and anisotropic interactions enhance polaron anisotropy. The isotope effect 
on the polaron mass and spectrum is discussed. A path-integral approach to the Jahn-Teller 
polaron is developed. Extensions of the method to lattice bipolarons and to more complex 
polaron models are outlined. 

\end{abstract}

\section{ \label{sec:one}
Introduction
}

The polaron problem was one of the first applications of path integrals (PI). Just a few years after 
the introduction of the new technique into quantum mechanics \cite{Dirac1933,Feynman1948} and quantum 
statistics
\cite{Kac1949} Feynman published his seminal paper \cite{Feynman1955} on the theory of Fr\"ohlich 
polaron \cite{Froehlich}, which set a stage for the development of polaron physics for the next half 
a century. The main feature of the statistical PI is an extra dimension that transforms each point-like
particle into a one-dimensional line, or {\em path}. The extra dimension, to be called here the 
imaginary time $\tau$, extends between $0$ and $\beta = (k_B T)^{-1}$ where $T$ is the absolute 
temperature at equilibrium. As a result, a quantum-mechanical system is mapped onto a purely classical 
system in one extra dimension. This enables visualization of quantum-mechanical objects, which is an 
instructive and useful feature. The quantum mechanical properties are fully recovered by considering 
an ensemble of paths, each contributing its own statistical weight into the system's partition function.   
     
Having formulated the polaron problem in PI terms, Feynman made another key advance. He showed that 
if the ionic coordinates entered the model only quadratically (a free phonon field) and linearly 
(electron-phonon interaction in linear approximation) then the infinite-dimensional integration over 
ionic configurations could be performed analytically and exactly. As a result, {\em all}\/ bosonic 
degrees of freedom (an infinite number of them) are eliminated in favour of just one fermionic 
degree of freedom. Phonons remain in the system as a retarded {\em self-interaction} of the electron.  
In the PI language, the statistical weight of each path is exponential in its Euclidian action and the 
latter contains a double integral over imaginary time as opposed to a single integral in cases of 
ordinary instantaneous interactions. Different segments of the fermion path ``feel'' each other if 
they are separated in imaginary time by less than the inverse phonon frequency. As a result, the 
electron path stiffens which leads to an enhanced effective mass and other characteristic effects 
as detailed later in the chapter.
      
The resulting electron path integral could not be calculated analytically due to the complex nature
of the retarded self-interaction. Feynman resolved the difficulty by employing an original variational
principle, in which the exact polaron action was replaced by an approximate, but exactly solvable, 
quadratic action. That led to a remarkably accurate top bound for the energy of the Fr\"ohlich polaron
(see \cite{Devreese1966,Prokofev1998,Mishchenko2000}), which was only marginally improved by subsequent 
generalizations of the method \cite{Osaka1959,Abe,Saitoh1980,Fedyanin1982,DeFilippis2003}. 
The PI method became a workhorse of the Fr\"ohlich polaron research for decades and inspired numerous 
extensions which included polaron mobility \cite{Feynman1962,Thornber1970,Peeters1984} and optical 
conductivity \cite{Devreese2006}, polaron in a magnetic 
field \cite{Peeters1982,Devreese1992}, large bipolaron \cite{Verbist1990}, and others. Feynman's 
method also inspired a Fourier Monte Carlo method, in which only the {\em difference} between the 
exact and variational energy was calculated numerically \cite{Alexandrou1990,Alexandrou1992}. 
Recently, the variational PI treatment was extended to a many-polaron system \cite{Brosens2005}.

In the parallel development of the small-polaron physics, Feynman's remarkable reduction had not been 
utilized for almost thirty years. Although the phonon integration could be performed as well the 
self-interacting electron PI could not be calculated. The problem was that on the lattice even a 
free particle possessed a non-quadratic action. Because such a path integral could not be done 
analytically, it could not serve as a trial variational action. The situation changed in 1982 when 
De Raedt and Lagendijk (DRL) observed \cite{DeRaedt1982} that the electron PI could instead be 
evaluated numerically. Metropolis Monte Carlo \cite{Metropolis1953} was ideally suited for the task 
because the polaron action was purely real, and as such resulted in a positive-definite statistical 
weight of the path. Using this approach, DRL obtained a number of nice results on the Holstein 
polaron: confirmed formation of a self-trapped state with increasing strength of the electron-phonon 
interaction, observed that the crossover gets sharper with decreasing phonon frequency and 
increasing lattice dimensionality \cite{DeRaedt1982}, and that the critical coupling goes down
as a dispersive phonon mode softens \cite{DeRaedt1984}. These results were reviewed in 
\cite{DeRaedt1985}. DRL also provided the first Monte Carlo analysis of the Holstein bipolaron 
\cite{DeRaedtBipolaron}.    

The works of DRL were an important step forward in the application of PI to lattice polarons. 
Their method could handle infinite crystal lattices of arbitrary symmetry and dimensionality, 
dispersive phonons, and long-range electron-phonon interactions. At the same time, it was still 
limited to thermodynamical properties such as energy, specific heat, and static correlation 
functions. In addition, it suffered from a systematic error caused by finite discretization of 
imaginary time (the Trotter slicing). These limitations were removed in 1997-1998. Firstly, it 
was shown how open boundary conditions in imaginary time could enable direct calculation of the 
polaron effective mass \cite{KornilovitchPike1997} and even the entire spectrum and density of 
states \cite{Kornilovitch1999}. The open boundary conditions allow projection of the partition 
function on states with definite polaron momenta. This is a particular manifestation of the 
general projection technique in the presence of a global symmetry. Systematic application of 
this principle makes it possible to separate states of different symmetries in many interesting 
situations, most notably bipolarons of different parity and orbital symmetry. More about the 
projection technique is in Section~\ref{sec:two}. Secondly, the polaron Monte Carlo method was 
formulated in continuous imaginary time \cite{Kornilovitch1998}, which completely eliminated the 
Trotter slicing and tremendously improved the computational efficiency of the method. This will 
be described in detail in Section~\ref{sec:three}. As a result, the versatility of DRL's method 
was enhanced by better computational efficiency and by a number of new polaron and bipolaron 
properties that could be computed with path integrals. The method was further developed in
\cite{longrange,anisotropy,Kornilovitch2000,isotope,Spenser2005,Hague2005,multiphonon,Hague2006}, 
the content of which will be described later in the Chapter.

The path-integral Quantum Monte Carlo (PIQMC) with phonon integration is only one from an impressive 
list of numerical methods developed for polaron models in the last three decades. There exist 
several other QMC techniques, in particular the path-integral QMC without phonon integration 
\cite{Hirsch1982,HirschSSH}, Fourier QMC \cite{Alexandrou1990,Alexandrou1992}, diagrammatic QMC 
\cite{Prokofev1998,Mishchenko2000,Mishchenko2003,MishchenkoNagaosa2004,Macridin2004,Mishchenko2006} 
and Lang-Firsov QMC \cite{Hohenadler2004,Hohenadler2005,Hohenadler2006}. 
Non-QMC {\em classes} of methods include exact diagonalization 
\cite{Kabanov1994,Wellein1996,Wellein1997,Fehske2000,Bronold2002,Fehske2006}, 
variational calculations 
\cite{DeFilippis2003,Romero1998,Bonca1999,Bonca2000,Bonca2001,Ku2002,ElShawish2003,Cataudella2000,Perroni2004,Cataudella2006} 
and the density-matrix renormalization group \cite{Jeckelmann1998,Jeckelmann1999,Zhang1999}. 
Despite proliferation of methods, most of them have been applied to the two major polaron 
models: the ionic crystal model of Fr\"ohlich \cite{Froehlich} and molecular crystal model of 
Holstein \cite{Holstein1959}. (The notable exceptions are the Jahn-Teller (bi)polaron 
\cite{Kornilovitch2000,ElShawish2003} and Peierls instabilities 
\cite{HirschSSH,Wellein1998,Weisse1998}.) 
Due to its versatility, PIQMC is uniquely positioned to study many other models, for example 
long-range and anisotropic electron-phonon interactions. In addition, it provides some (bi)polaron
properties that are difficult to obtain by other methods such as the density of states. In this 
way, several physically interesting results have been obtained that include, in particular, the 
light polaron mass in the case of long-range electron-phonon interactions \cite{longrange}, 
the enhancement of polaron anisotropy by electron-phonon interaction \cite{anisotropy}, formation
of a peak in the polaron density of states \cite{Kornilovitch1999,Kornilovitch2000,Hague2005},
the isotope effect on the polaron spectrum and density of states \cite{isotope}, and the 
``superlight'' crab-like bipolaron \cite{Alexandrov1996,Kornilovitch2002,Hague2006}. These and 
other results obtained by the PIQMC method in the last decade will be reviewed in 
Section~\ref{sec:four}. Various extensions of PIQMC and its prospects are given in 
Section~\ref{sec:six}.

\section{ \label{sec:two}
Projected partition functions
}

It is commonly believed that the statistical path integral can provide information only on the 
ground state of a quantum mechanical system, and in general this is true. However, when the system 
possesses a global symmetry suitable projection operators can project the configurations on  
sectors of the Hilbert space corresponding to different irreducible representations of the 
symmetry group. That enables access to the lowest states within each sector and provides valuable 
information about system's excitations. This strategy proves very useful in the path-integral 
studies of polaron models. It enables calculation of the polaron mass, spectrum, density 
of states, bipolaron singlet-triplet splitting and other properties. The above considerations are 
formalized as follows. The full thermodynamic partition function is a trace of the density matrix:
\begin{equation}
Z = \sum_{{\bf R}} \langle {\bf R} \vert e^{-\beta H} \vert {\bf R} \rangle \: ,
\label{eq:one}   
\end{equation}
where $H$ is the Hamiltonian and $\vert {\bf R} \rangle$ is a complete set of basis states in the 
real-space representation. If there is a global symmetry group $G$ with a set of irreducible
representations $U$, it is meaningful to compose a projected partition function $G_U$ which is a 
trace over the $U$-sector of the Hilbert space only:
\begin{equation}
Z_U = \sum_{{\bf R}_U} \langle {\bf R}_U \vert e^{-\beta H} \vert {\bf R}_U \rangle 
    = \sum_{\bf R} \langle {\bf R} \vert O^{\dagger}_U e^{-\beta H} O_U \vert {\bf R} \rangle \: ,
\label{eq:two}   
\end{equation}
where operator $O_U$ generates a basis state $\vert {\bf R}_U \rangle$ from an arbitrary state 
$\vert {\bf R} \rangle$. In the low-temperature limit $\beta \rightarrow \infty$, the partition function 
is dominated by the lowest $U$-eigenvalue, $Z_U \rightarrow \exp{(-\beta E_U)}$. Thus $E_U$ can be found
by taking the logarithm of $Z_U$ in the low-temperature limit. This is particularly efficient if the
first excited state in $U$ is separated from $E_U$ by a finite gap. 
 
The most important application of this technique is the formula for the (bi)polaron effective mass.
In this case, irreducible representations $U$ are labeled by the total momentum ${\bf K}$ and the 
projection operator is $O_{\bf K} = \sum_{\bf r} e^{i {\bf K} {\bf r}} T_{\bf r}$, where $T_{\bf r}$ 
is the shift operator by a lattice vector ${\bf r}$. The ${\bf K}$-projected partition function is 
then \cite{Kornilovitch1999}
\begin{equation}
Z_{\bf K} = \sum_{\Delta {\bf r}} e^{i {\bf K} \Delta {\bf r}}
\langle {\bf R} + \Delta {\bf r} \vert e^{-\beta H} \vert {\bf R} \rangle 
          = \sum_{\Delta {\bf r}} e^{i {\bf K} \Delta {\bf r}} Z_{\Delta {\bf r}} \: .
\label{eq:three}   
\end{equation}
Here ${\bf R} + \Delta {\bf r}$ denotes a many-particle configuration ${\bf R}$, which is 
parallel-shifted as a whole by a lattice vector $\Delta {\bf r}$. The partition function 
$Z_{\Delta {\bf r}}$ is a ``shifted trace'' of the density matrix: it connects ${\bf R}$ not 
with ${\bf R}$ but with ${\bf R}$ shifted by $\Delta {\bf r}$. It follows from the above equation 
that $Z_{\bf K}$ and $Z_{\Delta {\bf r}}$ satisfy a Fourier-type relationship (for a more detailed 
derivation, cf. Refs.~\cite{Kornilovitch1999,Kornilovitch2005}). The partition function 
$Z_{\Delta {\bf r}}$ is formulated in real space and as such can be represented by a 
real-space path integral. The partition function $Z_{\bf K}$ is diagonal in momentum space 
and therefore contains information about the variation of system's properties with ${\bf K}$. 
The Fourier theorem (\ref{eq:three}) is key to the ability to infer the (bi)polaron spectrum 
and mass from the path integral.  

Dividing $Z_{\bf K}$ by $Z_{{\bf K} = 0}$ one obtains in the zero temperature limit
\begin{equation}
\frac{Z_{\bf K}}{Z_{{\bf K} = 0}} = e^{-\beta ( E_{\bf K} - E_0 )} =
\frac{ \sum_{\Delta {\bf r}} e^{i {\bf K} \Delta {\bf r}}
       \langle {\bf R} + \Delta {\bf r} \vert e^{-\beta H} \vert {\bf R} \rangle }
{ \sum_{\Delta {\bf r}} \langle {\bf R} + \Delta {\bf r} \vert e^{-\beta H} \vert {\bf R} \rangle }       
 = \langle \cos {{\bf K} \Delta {\bf r}} \rangle_{\rm shift} \: .
\label{eq:four}   
\end{equation}
The ratio of two sums over $\Delta {\bf r}$ represents the mean value of 
$\exp{( i {\bf K} \Delta {\bf r} )}$ over an ensemble of paths in which the two end-point many-body 
configurations are identical up to an arbitrary parallel shift. In the QMC language, simulations
need to be performed with {\em open boundary conditions in imaginary time}. There is a certain
parallel with the uncertainty principle: fixing the boundary conditions in imaginary time (by
making them periodic) results in a mix of all possible momenta. Conversely, relaxing the boundary 
conditions by mixing in all possible shifts $\Delta {\bf r}$ allows extraction of information on 
the given ${\bf K}$-sector of the Hilbert space. Resolving the last equation with respect to 
$E_{\bf K}$, one obtains an estimator for the (bi)polaron spectrum
\begin{equation}
E_{\bf K} - E_0 = - \lim_{\beta \rightarrow \infty} \frac{1}{\beta} 
\ln \langle \cos {{\bf K} \Delta {\bf r}} \rangle_{\rm shift} \: .
\label{eq:five}   
\end{equation}
The left-hand-side of this equation is a constant that depends on the parameters of the model
being simulated. Therefore, the average cosine on the opposite side must decrease exponentially 
with $\beta$ to compensate the growing denominator. At some $\beta$ it becomes too small to be 
measured reliably with the available statistics of a Monte Carlo run. This is an incarnation of 
the infamous ``sign problem'' that plagues many QMC algorithms. In physical terms, at very low
temperatures the probability to access an excited state is exponentially low due to the Boltzmann
weight factor. As a result, the QMC process samples relevant configurations exponentially rarely. 
These considerations put the low limit on the temperature of the simulation (high limit on $\beta$).
At the same time, the condition that $Z_{\bf K}$ is dominated by a single eigenstate in the 
${\bf K}$ sector, puts a high limit on temperature. Thus there is an interval of temperatures
at which a meaningful calculation of (bi)polaron spectrum can be performed. If the expected 
bandwidth is too large the temperature interval shrinks to zero, which renders the calculation 
impossible. In general, the (bi)polaron spectrum can be computed by this method if the total 
bandwidth is smaller than the phonon frequency.

Expanding (\ref{eq:five}) for small momenta, one derives an estimator for the $\mu$-th 
component of the {\em inverse} effective mass
\begin{equation}
\frac{1}{m^{\ast}_{\mu}} = \lim_{\beta \rightarrow \infty} 
\frac{\langle ( \Delta r_{\mu} )^2 \rangle_{\rm shift}}{\hbar^2 \beta}  \: .
\label{eq:six}   
\end{equation}
Since in the present formulation the end points of the path are not tied together path evolution  
between $\tau = 0$ and $\tau = \beta$ can be regarded as {\em diffusion} of the top end with 
respect to the bottom end over a diffusion time $\beta$. According to the Einstein relation, the 
mean square displacement of a diffusing particle is proportional to time. Then (\ref{eq:six}) 
implies that the {\em inverse effective mass is proportional to the diffusion coefficient $D$ of 
the path in imaginary time} \cite{KornilovitchPike1997,Kornilovitch1998,Kornilovitch2005}. The 
combination $\hbar \beta$ has the dimensionality of time, so the inverse mass can be written as
\begin{equation}
\frac{1}{m^{\ast}_{\mu}} = \frac{2D}{\hbar}  \: .
\label{eq:seven}   
\end{equation}
This formula works for both polarons and bipolarons \cite{Macridin2004,Hague2006}. It can be viewed 
as a fluctuation-dissipation relation because it equates a dynamical characteristic, the mass, 
with an equilibrium thermodynamic property $D$. In addition, it provides a nice visualization of 
the main polaron effect: effective mass increase caused by an electron-phonon interaction. As will be 
shown in the next section, phonon integration results in correlations between distant parts of the 
imaginary-time polaron path. This increases the statistical weight of paths with straight segments. 
The average polaron path {\em stiffens}, which translates in a reduced diffusion coefficient $D$ 
and, according to (\ref{eq:seven}), enhanced effective mass. Note that these considerations apply 
equally well to other types of composite particles whose mass originates from interaction. The most 
notable examples are defects in quantum liquids \cite{Boninsegni1995} and the hadrons of quantum 
chromodynamics. 

Another important application of the projection technique is the singlet-triplet splitting of the 
bipolaron, or, more generally, of a two-fermion bound state. Coming back to the projected partition
function $Z_U$, (\ref{eq:two}), the relevant projection operators are identified as $O_S = I + X$ 
for singlet states and $O_T = I - X$ for triplet states. Here $I$ is the identity operator and $X$ is 
the fermion exchange operator, 
$X \vert {\bf r}_1 {\bf r}_2 \{ \xi \} \rangle = \vert {\bf r}_2 {\bf r}_1 \{ \xi \} \rangle$, where
$\{ \xi \}$ are the ionic displacements. Upon substitution in (\ref{eq:two}) one finds
\begin{equation}
Z^{S,T} = \sum_{{\bf r}_1 {\bf r}_2 \{ \xi \}} 
\langle {\bf r}_1 {\bf r}_2 \{ \xi \} \vert e^{-\beta H} 
\vert {\bf r}_1 {\bf r}_2 \{ \xi \} \rangle
\pm \langle {\bf r}_2 {\bf r}_1 \{ \xi \} \vert e^{-\beta H} 
\vert {\bf r}_1 {\bf r}_2 \{ \xi \} \rangle  \: .
\label{eq:eight}   
\end{equation}
Monte Carlo simulation proceeds over an ensemble of paths that have identical end configurations 
{\em up to exchange of the top ends of the two fermion trajectories}. For the singlet, both types of
configurations contribute a phase factor $(+1)$ whereas for the triplet the direct paths contribute
$(+1)$ and exchanged paths contribute $(-1)$. Forming the ratio of the triplet and singlet partition 
functions one obtains an estimator for the $S-T$ energy split
\begin{equation}
E^{T}_{0} - E^{S}_{0} = - \lim_{\beta \rightarrow \infty} \frac{1}{\beta} 
\ln \langle (-1)^P \rangle_{\rm ex} \: .
\label{eq:nine}   
\end{equation}
Here $(-1)^P = \pm 1$ for the direct and exchanged two-fermion paths. If the ends of the two paths
coincide then $(-1)^P = 0$. All said above about the sign problem and the role of temperature in 
calculating the average cosine applies to this expression as well. 

Equation (\ref{eq:nine}) provides the energy difference between the lowest triplet and singlet
states. It is possible to compute the effective mass and the entire spectrum of the triplet bipolaron.
To this end, one needs to combine the two symmetries considered above: the translational and exchange.
This naturally leads to a path ensemble with the end configurations differing by any combination of
parallel shifts and exchanges. Without repeating the derivation, the final expressions are
\begin{equation}
E^T_{\bf K} - E^{T}_{0} = 
- \frac{1}{\beta} \ln \left( \frac{Z^T_{\bf K}}{Z^S_0} \frac{Z^S_0}{ Z^T_{0}} \right) =
- \frac{1}{\beta} \ln \frac{\langle (-1)^P \cos ({\bf K} \Delta {\bf r}) \rangle_{\rm shift,ex}}
{\langle (-1)^P \rangle_{\rm shift,ex}}  \: ,
\label{eq:ten}
\end{equation} 
for the triplet spectrum and 
\begin{equation}
\frac{1}{m^T_i} = \frac{1}{\beta \hbar^2} 
\frac{\langle (-1)^P ( \Delta r_i )^2 \rangle_{\rm shift,ex}}
{\langle (-1)^P \rangle_{\rm shift,ex}}  \: ,
\label{eq:eleven}
\end{equation} 
for the triplet mass. For the singlet spectrum and mass, (\ref{eq:five}) and (\ref{eq:six})
are still valid, except that the boundary conditions must be changed to ``shift and 
exchange''. The same applies to the singlet-triplet estimator (\ref{eq:nine}).

\section{ \label{sec:three}
Continuous-time path-integral quantum Monte Carlo method
}

The main appeal of quantum Monte Carlo (QMC) methods is the ability to calculate physical properties 
without approximations. A quantum mechanical system is directly {\em simulated} taking into account
all the details of particle dynamics and inter-particle interaction. Physically interesting 
observables can be calculated without bias while the statistical errors in general decrease with 
increasing simulation time. It is said therefore that QMC provides ``numerically exact''
values for the observables. A number of comprehensive reviews on QMC exist 
\cite{Binder1978,Binder1984,DeRaedt1985,Binder1995,Ceperley1995,Foulkes2001}. Many problems of the 
early QMC methods, such as the finite-size effects, finite time-step effects, and critical slowing 
down, have been resolved with the development of novel algorithms and increasing computing power. 
The {\em sign problem}, that is the non-positive-definiteness of the statistical weight of
basis configurations, remains the only fundamental problem. In real systems and models without 
a sign problem (for example liquid and solid He$^4$ 
\cite{Ceperley1995,Gordillo1998,Bauer2000,Ceperley2004,Clark2006}), QMC works beautifully and 
provides with accurate and valuable information about the thermodynamics and sometimes real-time 
dynamics.   
  
{\em One} polaron is free from the sign-problem difficulties. Phonons are bosons and as such do not
lead to a fermion sign problem. And as long as there is only one polaron in the system statistics 
does not matter. This is the main reason behind the success of the QMC approach to the polaron.
The ground state energy, the effective mass, isotope exponents on mass, the number of excited phonons, 
static correlation functions and other quantities can be calculated without any approximations. 
The bipolaron ground state can also be investigated without a sign problem as long as the bipolaron 
is a spin singlet with a symmetric spatial wave function. Thus the power of QMC can be (and has been)
applied to bipolarons of various kinds and to pairs of distinguishable particles such as the exciton. 
For many-polaron systems the fermion sign problem fully manifests itself, which limits
the applicability of QMC methods. More about this will be said in Section~\ref{sec:six}. In this 
Section the basics of the continuous-time path-integral quantum Monte Carlo (PIQMC) method for 
lattice polarons are described. 

The starting point is the shift partition function $Z_{\Delta {\bf r}}$, see (\ref{eq:three}), 
and the electron-phonon (e-ph) Hamiltonian   
\begin{equation}
H = - t \sum_{\bf n n'} c^{\dagger}_{\bf n'} c_{\bf n} - 
\sum_{\bf nm} f_{\bf m}({\bf n}) c^{\dagger}_{\bf n'} c_{\bf n} \xi_{\bf m} +
\sum_{\bf m} \left( - \frac{\hbar^2}{2M} \frac{\partial^2}{\partial \xi^2_{\bf m}} + 
\frac{M \omega^2}{2} \, \xi^2_{\bf m} \right) \: .
\label{eq:fifteen}
\end{equation} 
The Hamiltonian is written in mixed representation, $c_{\bf n}$ being fermionic 
annihilation operators and $\xi_{\bf m}$ ion displacements. Index ${\bf n}$ denotes the spatial
location of an electron Wannier orbital whereas ${\bf m}$ denotes that of an ion
displacement. In general, ${\bf n} \neq {\bf m}$ even if they belong to the same lattice unit cell.
For simplicity, the electron kinetic energy (the first term of $H$) is taken in the 
nearest-neighbor-hopping approximation with amplitude $-t$, and the lattice energy (third term
in $H$) in the independent Einstein oscillator approximation with mass $M$ and frequency $\omega$.
Neither approximation is necessary. Section~\ref{sec:six} explains how PIQMC deals with 
more complex forms of kinetic energy and phonon spectrum. The most interesting part is the 
e-ph interaction (second term in $H$) which is written in the ``density-displacement'' form.   
The quantity $f_{\bf m}({\bf n})$ is the {\em force} with which an electron ${\bf n}$ acts on
the ion coordinate ${\bf m}$. Asymmetric notation emphasizes that
the force $f_{\bf m}$ is an attribute of a given oscillator, while the argument ${\bf n}$ is
a dynamic variable indicating the current position of the electrons interacting with ${\bf m}$. 
No constraints are imposed on the functional form of $f$, thus allowing 
studies of long-range e-ph interactions. The commonly used Holstein model \cite{Holstein1959} 
corresponds to a localized force function $f_{\bf m}({\bf n}) = \kappa \delta_{\bf nm}$.

\subsection{ \label{sec:threeone}
Handling kinetic energy in continuous time
}

Handling the kinetic energy on a lattice possesses a challenge for Monte Carlo. To see this consider
just the kinetic part of $H$ and one free particle. By introducing multiple resolutions of identity,
$\hat{I} = \sum_{\bf r} \vert {\bf r} \rangle \langle {\bf r} \vert$, the shift partition function is
developed into a multidimensional sum
\begin{equation}
Z_{\Delta {\bf r}} = \sum_{{\bf r}_1 {\bf r}_2 \ldots {\bf r}_M} 
\langle {\bf r}_1 + \Delta {\bf r} \vert e^{- \Delta \tau H_{\rm el}} \vert {\bf r}_M \rangle 
\ldots \langle {\bf r}_3 \vert e^{- \Delta \tau H_{\rm el}} \vert {\bf r}_2 \rangle 
       \langle {\bf r}_2 \vert e^{- \Delta \tau H_{\rm el}} \vert {\bf r}_1 \rangle \: ,
\label{eq:twentyone}
\end{equation} 
where $\Delta \tau = \beta/L$ is the time step and $L+1$ is the number of time slices. Each 
matrix element from the product is expanded to the first power in $\Delta \tau$:
\begin{equation}
  \langle {\bf r}_{j+1} \vert e^{- \Delta \tau H_{\rm el}} \vert {\bf r}_j \rangle
= \delta_{{\bf r}_{j+1}, {\bf r}_j} + t \Delta \tau  \sum_{\bf l}  
\delta_{{\bf r}_{j+1}, \, {\bf r}_j  + {\bf l}}  
+ {\cal O}(\Delta \tau ^2 ) \: ,
\label{eq:twentytwo}
\end{equation} 
where ${\bf l}$ runs over the nearest neighbors. Expanding the product, the partition function 
becomes a sum of a large number of terms, each of which represents a particular path of the electron 
in imaginary time ${\bf r}(\tau)$. The paths consist of two different building blocks: straight 
segments and kinks, which originate from the first and second terms in (\ref{eq:twentytwo}), 
respectively. On straight segments the electron position does not change, whereas each kink 
changes the electron coordinates by vector ${\bf l}$ depending on the kink type. The statistical
weight of a path with $N_k$ kinks is 
\begin{equation}
W_{N_k} = \underbrace{(t \Delta \tau) (t \Delta \tau) \ldots (t \Delta \tau)}_{N_k \; {\rm times}} 
= (t \Delta \tau)^{N_k} \: .
\label{eq:twentythree}
\end{equation} 
Now consider two paths, $D$ and $D'$, that differ by $D'$ having one extra kink of type ${\bf l}$ 
at imaginary time $\tau$. At small $\Delta \tau$, $D'$ has a vanishingly small statistical weight 
compared to $D$. As a result, no meaningful detailed balance between {\em individual} paths with 
different number of kinks seems to be possible in the continuous-time limit. The resolution of this 
difficulty comes from the observation that there are {\em infinitely many} paths with $N_k+1$ kinks 
than with $N_k$ kinks. The small weight of individual path is compensated by the large number of paths. 
The combined weight of all paths with $N_k+1$ kinks is of the same order than the combined weight
of all paths with $N_k$ kinks. Instead of detailed balance between individual paths one should seek 
detailed balance between path groups of similar combined weight. The paths $D$ and $D'$ are regarded
as representatives of such groups. 

In accordance with the general ideas of the Metropolis Monte Carlo 
\cite{Metropolis1953}, one has to compose 
an equation that equates the transition rate from $D$ to $D'$ to the reciprocal transition rate 
from $D'$ to $D$. Both rates are products of three factors: the probability to have the original 
path, which is proportional to the path's weight $W$, the probability $Q$ to propose the necessary 
change to the path, and the probability $P$ to accept the change. The detailed balance reads:
\begin{equation}
W(D)  Q({\rm add}    \; {\bf l} \; {\rm at} \; \tau) P(D  \rightarrow D') = 
W(D') Q({\rm remove} \; {\bf l} \; {\rm at} \; \tau) P(D' \rightarrow D )  \: .
\label{eq:twentyfour}
\end{equation} 
The probabilities to propose the addition and removal of the kink turn out to be two quite different
quantities. In the direction $D' \rightarrow D$, the kink to be removed is chosen from a {\em finite} 
number of all other kinks. Therefore the probability that it is selected for removal is a finite 
number $Q_{\rm remove}({\bf l}; \tau)$. In contrast, in going from $D$ to $D'$ the kink does not yet 
exist and the probability to propose its creation {\em exactly at time} $\tau$ is proportional to 
the time interval $\Delta \tau$. The probability to propose the kink's addition is an infinitesimal 
quantity: $Q({\rm add}\; {\bf l}\; {\rm at}\; \tau) = q_{\rm add}({\bf l}; \tau)(\Delta \tau)$, 
where $q_{\rm add}$ is the corresponding probability {\em density}. Substitution of these two results 
in the balance equation yields
\begin{equation}
(t \Delta \tau)^{N_k} q_{\rm add}({\bf l} ; \tau)(\Delta \tau) P(D  \rightarrow D') = 
(t \Delta \tau)^{N_k+1} Q_{\rm remove}({\bf l} ; \tau) P(D' \rightarrow D )  \: .
\label{eq:twentyfive}
\end{equation} 
The extra $(\Delta \tau)$ from the addition probability on the left side exactly 
balances the smaller weight of the right side. The time step cancels out of the equation leaving
only finite factors which are well defined in the continuous-time limit. According to the 
standard Metropolis recipe the solution is given by the following acceptance rules
\begin{equation}
P(D \rightarrow D') = \min \left\{ 1 \; ; \;
\frac{t \cdot Q_{\rm remove}({\bf l}; \tau)} {q_{\rm add}({\bf l}; \tau)}  \right\}  \: ,
\label{eq:twentysix}
\end{equation} 
\begin{equation}
P(D' \rightarrow D) = \min \left\{ 1 \; ; \; 
\frac{q_{\rm add}({\bf l}; \tau)} {t \cdot Q_{\rm remove}({\bf l}; \tau)} \right\}  \: .
\label{eq:twentyseven}
\end{equation} 
These expressions still leave a lot of freedom in specifying functions $q_{\rm add}$ and 
$Q_{\rm remove}$. As an example, consider the simplest choice when both functions are constant.
For the addition process this means that the time for the new kink of type ${\bf l}$, on top of
the existing $N_{\bf l}$, is chosen with equal probability within the $[0,\beta]$ time interval,
hence ${q_{\rm add}({\bf l}; \tau)} = \beta^{-1}$. For the removal process, the candidate kink for
elimination is chosen from the existing $N_{\bf l}$ with equal probability, that is 
$Q_{\rm remove}({\bf l}; \tau) = N^{-1}_{\bf l}$. One must also take into account the fact that   
kinks can only be added to the path if $N_{\bf l} = 0$, while added or removed if 
$N_{\bf l} \geq 1$. The full set of update rules could be as follows. (i) Select the kink type 
${\bf l}$, that is the hopping direction of the particle. The probability with which different 
${\bf l}$ are selected may not be equal but must be non-zero. (ii) If the current path already has 
kinks of type ${\bf l}$, whether to add or remove a kink is proposed with, for example, equal 
probability of $\frac{1}{2}$. If the path possesses no kinks, addition is the only option, and it 
must be chosen with probability $1$. (iii) If addition is selected, the time for the new kink is
chosen with probability density $\frac{1}{\beta}$. The update is accepted with probability 
$\min \{ 1 ; (t\beta)/(N_{\bf l}+1) \}$ for $N_{\bf l} \geq 1$ and with probability 
$\min \{ 1 ; (t\beta)/2 \}$ for $N_{\bf l} = 1$. (iv) If removal is selected, the candidate kink
is chosen with equal probability from the existing $N_{\bf l}$. The update is accepted with 
probability $\min \{ 1 ; N_{\bf l}/(t\beta) \}$ for $N_{\bf l} \geq 2$ and with probability
$\min \{ 1 ; 2/(t\beta) \}$ for $N_{\bf l} = 1$.
   
The described method can be readily generalized to next-nearest neighbor hopping and anisotropic
hopping. It can also be generalized to multi-kink updates, as needed for example in simulating the
bipolaron. To close this section one should note that a path can be interpreted as a time-space 
{\em diagram}, where kinks play the role of vertices 
and straight segments the role of propagators. Integration over the ensemble of fluctuating paths 
can then be understood within the general concept of Diagrammatic Monte Carlo
\cite{Prokofev1998,ProkofevDMC}.

\subsection{ \label{sec:threetwo}
Integration over phonons
}

Analytic integration over phonons is a key ingredient of the polaron PIQMC method. Although an 
electron-phonon (e-ph) system can be simulated without phonon integration 
\cite{Hirsch1982,HirschSSH}, it is the integration that makes the method powerful. Indeed, the 
integration reduces $N+1$ degrees of freedom to just one degree of freedom, which increases 
accuracy and reduces simulation time. The integration effectively eliminates all the distant 
parts of the system by folding the infinite number of ion coordinates into one retarded 
self-interaction function. The simulation then proceeds on an infinite lattice with no 
finite-size effects. In addition, the integration makes it possible to simulate long-range 
e-ph interactions with the same efficiency as local interactions, which extends QMC 
capabilities towards more realistic models.

Technical details of the phonon integration are somewhat cumbersome but well-documented in the
literature. Here we will describe the main steps of the derivation and point out associated 
subtleties. 

\begin{figure}
\centering
\includegraphics[width=11.7cm]{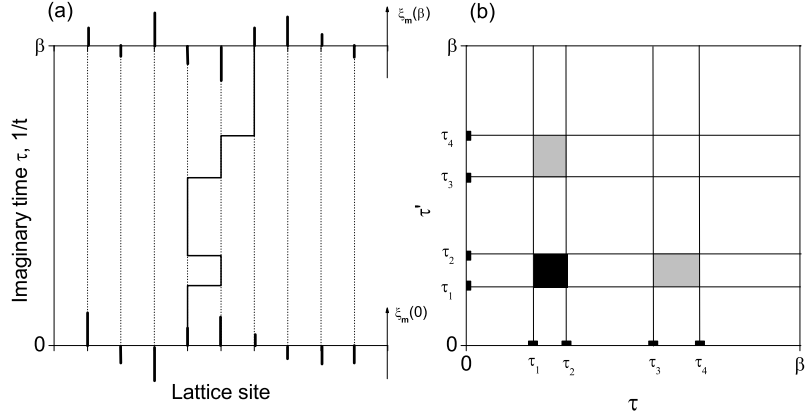}
\vspace{-0.5cm}
\caption{
(a) Integration over ionic paths under shifted boundary conditions. Ionic displacements are shown 
as vertical bars at $\tau = 0$ and $\tau = \beta$. Notice how the pattern of displacements is 
correlated with the shift of the electron path. The end displacements of individual oscillators 
are not equal, $\xi_{\bf m}(0) \neq \xi_{\bf m}(\beta)$. 
(b) Calculation of the polaron action as a double integral over imaginary time. The kink
times $\tau_i$ break the $(\tau \tau')$ plane into a finite number of rectangles. Within
each rectangle the electron coordinates are constant and the double integral can be 
calculated analytically for arbitrary $\tau_i$ \cite{Spenser2005}. After that the double
integral reduces to a sum over the rectangles. Since the number of rectangles decreases
at strong coupling, the algorithm gets faster at {\em strong} coupling. 
}
\label{fig:two}       
\end{figure}

Upon insertion of $H$ into (\ref{eq:three}), a path-integral representation for 
$Z_{\Delta {\bf r}}$ is developed by introducing $L$ time intervals and $L-1$ intermediate
sets of basis states. In the continuous time limit, $L \rightarrow \infty$, the kinetic energy 
of the electron is treated as described in the previous section. Integration over the electron 
coordinates is converted to stochastic summation over an ensemble of paths which are sampled by 
adding and removing kinks. For each electron path, there is an internal path integration over 
the ionic coordinates characterized by the electron-phonon action $A_{\rm e-ph}$ 
\begin{equation}
Z_{\Delta {\bf r}} = \int^{({\bf r} + \Delta {\bf r}, \beta)}_{({\bf r},0)} {\cal D} {\bf r}(\tau) 
\int^{ ( \{ \xi_{{\bf m}-\Delta {\bf r}} \} , \beta ) }_{(\{ \xi_{\bf m} \} , 0)} 
{\cal D} \xi_{\bf m}(\tau) \: e^{A_{\rm e-ph} [ \, {\bf r}(\tau), \, \xi_{\bf m}(\tau)]} \: ,
\label{eq:twentyeight}
\end{equation}
\begin{equation}
A_{\rm e-ph} = - \sum_{\bf m} \!\! \int^{\beta}_{0} \! 
\left[ \frac{M {\dot \xi}^2_{\bf m}(\tau)}{2 \hbar^2} + \frac{M\omega^2}{2} \xi^2_{\bf m}(\tau)
\right] {\rm d} \tau + \sum_{\bf m} 
\int^{\beta}_{0} \!\!\! f_{\bf m}[{\bf r}(\tau)] \xi_{\bf m}(\tau) {\rm d} \tau  \: .
\label{eq:twentynine}
\end{equation}
The path integral over $\xi_{\bf m}(\tau)$ is Gaussian and therefore can be calculated analytically. 
Actually, the integration is performed in two steps. Firstly, {\em path} integration is done with 
fixed but arbitrary boundary conditions at the two ends of the path. Secondly, a certain 
relationship between $\xi_{\bf m}(0)$ and $\xi_{\bf m}(\beta)$ is assumed followed by an additional 
one-dimensional {\em ordinary} (but still Gaussian) integration. In most PI studies, 
periodic boundary conditions $\xi_{\bf m}(\beta) = \xi_{\bf m}(0)$ are assumed. This leads to the 
classic Feynman formula for the oscillator action in the presence of an arbitrary time-dependent 
external force \cite{Feynman1972}. However, we have seen in the previous section that mass and 
spectrum calculation requires a partition function with the end configurations parallel-shifted 
with respect to each other. The shift of the ionic configuration must be the same as that of the 
fermion configuration, which implies $\xi_{\bf m}(\beta) = \xi_{{\bf m} - \Delta {\bf r}}(0)$, as 
indicated in the last equation. The correlation between the electron and phonon boundary conditions 
is illustrated in Fig.~\ref{fig:two}(a).

Due to an additional shift, the second-phase integration results in a correction to the 
conventional ``periodic'' action. The full polaron action, upon summation over all oscillators, 
reads \cite{Kornilovitch1998,multiphonon} 
\begin{equation}
A_{\rm pol}[{\bf r}(\tau)] = A_0 + \Delta A \: ,
\label{eq:sixteen}
\end{equation}
\begin{equation}
A_{0} = \sum_{\bf m} \frac{\hbar}{4\omega M}
\int^{\beta}_0 \!\!\! \int^{\beta}_0 {\rm d} \tau {\rm d} \tau' 
\frac{\cosh \hbar \omega ( \frac{\beta}{2} - \vert \tau - \tau' \vert ) } 
{\sinh \hbar \omega \frac{\beta}{2}} f_{\bf m}[{\bf r}(\tau)] f_{\bf m}[{\bf r}(\tau')]  \: ,
\label{eq:seventeen}
\end{equation}
\begin{equation}
\Delta A = \sum_{\bf m} \frac{\hbar}{2\omega M} B_{\bf m} 
\left( C_{{\bf m} + \Delta {\bf r}} - C_{\bf m} \right)  \: ,
\label{eq:eighteen}
\end{equation}
\begin{equation}
B_{\bf m} \equiv
\int^{\beta}_0 {\rm d} \tau' \frac{\sinh \hbar \omega (\beta-\tau')}
                     {\sinh \hbar \omega \beta} f_{\bf m}[{\bf r}(\tau')]  \: , 
\label{eq:nineteen}
\end{equation}
\vspace{-0.2cm}
\begin{equation}
C_{\bf m} \equiv
\int^{\beta}_0 {\rm d} \tau' \frac{\sinh \hbar \omega \tau'}
                     {\sinh \hbar \omega \beta} f_{\bf m}[{\bf r}(\tau')]  \: . 
\label{eq:twenty}
\end{equation}
The polaron action is a functional of the electron path and contains all the information about
the ionic subsystem. The denominator of (\ref{eq:four}) finally becomes
\begin{equation}
Z_{{\bf K} = 0} = \sum_{\Delta \tau} Z_{\Delta {\bf r}} = Z_{\rm ph} 
\sum^{\infty}_{N_k = 0,1, \ldots} \int^{\beta}_{0} \cdots \int^{\beta}_{0} 
({\rm d}\tau)^{N_k} \, t^{N_k} \, e^{A_{\rm pol} [ \, {\bf r}(\tau)]} \: .
\label{eq:thirty}
\end{equation}
Here $Z_{\rm ph} = [2 \sinh (\frac{1}{2} \hbar \beta \omega) ]^{-N}$ ($N$ is the number of
lattice sites) is the partition function of free phonons. It is a multiplicative constant that 
cancels out from all statistical averages. The factor $e^{A_{\rm pol}}$ adds to the weight
of each path on top of the ``kinetic'' contribution $(t \Delta \tau)^{N_k}$. As a result, the
acceptance rules (\ref{eq:twentysix}) and (\ref{eq:twentyseven}) are modified by the ratio
of the respective factors for the paths $D$ and $D'$ 
\begin{equation}
P(D \rightarrow D') = \min \left\{ 1 \; ; \;
\frac{t \cdot Q_{\rm remove}({\bf l}; \tau)} {q_{\rm add}({\bf l}; \tau)}  
\: e^{A_{\rm pol}(D') - A_{\rm pol}(D) } \right\}  \: ,
\label{eq:thirtyone}
\end{equation} 
\begin{equation}
P(D' \rightarrow D) = \min \left\{ 1 \; ; \; 
\frac{q_{\rm add}({\bf l}; \tau)} {t \cdot Q_{\rm remove}({\bf l}; \tau)} 
\: e^{A_{\rm pol}(D) - A_{\rm pol}(D') } \right\}  \: .
\label{eq:thirtytwo}
\end{equation} 
\begin{figure}[t]
\centering
\includegraphics[width=11.7cm]{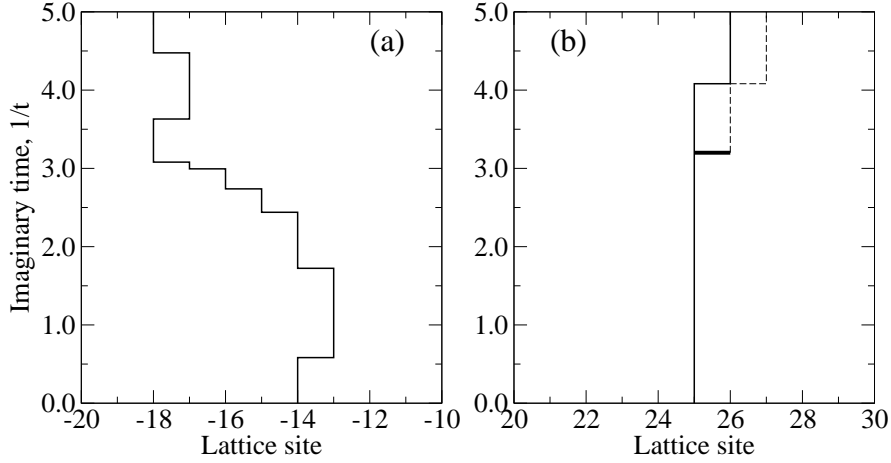}
\caption{
(a) A typical fluctuating path of a free particle on a one-dimensional chain ($E_0 = -2.00 \, t$,
$m^{\ast}/m_0 = 1$). The path has drifted to the left by four lattice sites, which contributes to 
the effective mass according to (\ref{eq:six}). (b) A typical path of a one-dimensional
Holstein polaron at $\hbar\omega/t = 2.0$ and $\lambda = 2.0$ ($E_0 = -4.66(2) \, t$, 
$m^{\ast}/m_0 = 3.06(3)$. The path has drifted to the right by just one site reflecting a larger
mass. The thick bar is a kink update proposed by Monte Carlo. If accepted, the top part of the 
path is changed to the dashed line ({\em top} shift). Alternatively, the bottom part of the path
can be shifted by one site to the left ({\em bottom} shift).   
}
\label{fig:one}       
\end{figure}

These acceptance rules together with the formula for the polaron action and path update rules 
described in the previous section fully define the Monte Carlo update process. It is illustrated 
in Fig.~\ref{fig:one}. Numerical efficiency of the algorithm critically depends on how easily the 
polaron action can be computed at every update. The task is greatly aided by the fact that the 
electron path is defined on a discrete lattice. In other words, it consists of a finite number of 
straight segments. This has two important consequences. First, because within each segment the 
electron coordinate is independent of imaginary time $\tau$, the forces $f_{\bf m}({\bf r})$ can be 
taken outside the time integration. The rest of the integrand is an explicit function of $\tau$ and 
$\tau'$. This time integral can be calculated analytically for arbitrary end times of the segment. 
The double integral over time becomes a double sum over the segments, as illustrated in 
Fig.~\ref{fig:two}(b). The summand is a simple function of kink times; the explicit expressions are 
given in \cite{Spenser2005}. The smaller the number of segments, that is the straighter the path, 
the less time is required to compute $A_{\rm pol}$. The electron-phonon interaction increases the
statistical weight of straight paths and reduces the mean number of segments. Thus, in contrast
to most other numerical methods, PIQMC becomes {\em faster} at large electron-phonon couplings. 

Secondly, the $(ij)$-th term of the sum over segments contains the electron-ion forces in
the combination
\begin{equation}
\Phi ( {\bf r}_i - {\bf r}_j )  = \sum_{\bf m} f_{\bf m}({\bf r}_i) f_{\bf m}({\bf r}_j) \: ,
\label{eq:thirtythree}
\end{equation} 
where ${\bf r}_i$ is the electron coordinate on $i$-th segment. The coefficients $\Phi$ are defined
on a discrete set of points and can be pre-computed for a sufficiently large range of coordinate 
separation. After that the simulation takes essentially the same amount of time for {\em any} form 
of electron-phonon interaction. This is why the PIQMC polaron method is as efficient for long-range 
e-ph interactions as for the short-range Holstein model. 

We close this section by noting that the requirement of phonon integration with shifted boundary 
conditions equally applies to the continuous case. This does not seem to be the case with the 
original Feynman's calculation of the mass of the Fr\"ohlich polaron and its subsequent 
generalizations. This issue was thoroughly investigated in \cite{KornilovitchFeynman2005}. It was 
shown that the periodic boundary conditions in the phonon integration, while absolutely correct 
for the energy calculation, lead to infinite terms in the polaron action in the mass calculation. 
Feynman's mass formula is obtained only if the terms are regarded as unphysical and thrown out by 
hand. In contrast, the shifted boundary conditions result in a self-consistent action and Feynman 
result is obtained without any complications.

\subsection{ \label{sec:threethree}
Observables
}

The preceding sections explained how to organize an efficient Monte Carlo sampling process that
simulates the electron kinetic energy while fully taking into account the ion dynamics and 
electron-phonon interaction. Note that {\em no} approximation has been made so far except for the
quadratic form of the elastic energy of the crystal. In this section it will be explained how
to extract useful physical information from the one-to-one correspondence between the quantum
mechanical-polaron and an ensemble of self-interacting paths. A number of important estimators
have already been derived in Section~\ref{sec:two}. In particular, the effective mass is obtained
from the mean-square end-to-end distribution, see (\ref{eq:six}). The polaron spectrum 
{\em relative to the ground state} is given by (\ref{eq:five}). It is essential that statistics 
for different $E_{\bf K}$ is collected in parallel so that the entire polaron spectrum is  
obtained in a single run. One consequence of this remarkable property is PIQMC's ability to 
compute the polaron density of states by simply histogramming $E_{\bf K}$ at the end. At present, 
PIQMC is the only method capable of efficiently calculating the polaron density of states.

The next important characteristic is the absolute polaron energy $E_0$. One way to obtain $E_0$
is from the low-temperature limit of the internal energy $U$. The latter can be computed from the 
thermodynamic relation $U = - \partial \ln Z /\partial \beta$. Since we are interested in the 
ground state, the corresponding partition function must be $Z_{{\bf K} = 0} = 
\sum_{\Delta {\bf r}} Z_{\Delta {\bf r}} = \sum_{\rm shifted \: paths} W[{\bf r}(\tau)]$.
Returning momentarily to discrete time, $\beta = L \Delta \tau$,  
\begin{equation}
U = - \frac{1}{L} \frac{\sum_{\rm shift} \frac{\partial W}{\partial \Delta \tau}}
{\sum_{\rm shift} W} = 
- \frac{1}{L} \frac{\sum_{\rm shift} 
\left[ \frac{1}{W} \frac{\partial W}{\partial \Delta \tau} \right] W } 
{\sum_{\rm shift} W} 
= - \frac{1}{M} \left\langle \frac{1}{W} \frac{\partial W}{\partial \Delta \tau} 
\right\rangle_{\rm shift} .
\label{eq:thirtyfour}
\end{equation} 
The weight $W$ is a product of the kinetic contribution $(t \Delta \tau)^{N_k}$ and the 
phonon term $e^{A_{\rm pol}}$. Substituting in the above expression, one obtains a 
ground-state energy estimator
\begin{equation}
E_0 = - \frac{1}{L} \left\langle \frac{N_k}{\Delta \tau} + 
\frac{\partial A_{\rm pol}}{\partial \Delta \tau} \right\rangle_{\rm shift} = 
\left\langle - \frac{N_k}{\beta} - \frac{\partial A_{\rm pol}}{\partial \beta} 
\right\rangle_{\rm shift} .
\label{eq:thirtyfive}
\end{equation} 
The two terms here represent the kinetic and potential energy of the polaron, respectively. 
The kinetic energy is simply the mean number of kinks on the path. One can regard the kinks
as ``quanta'' of the kinetic energy, each contributing the same amount $-\beta^{-1}$.
With increasing coupling the paths become ``stiffer'', that is the mean number of kinks
goes down. As a result, the polaron kinetic energy decreases by absolute value. Measuring
the potential energy is harder for it is derived from the polaron action. Similarly to the
latter, the potential energy estimator can be reduced to a double sum over the path's segments.
Explicit expressions for the double sum terms are given in \cite{Spenser2005}.

The number of excited phonons $N_{\rm ph}$ in the polaron cloud is another interesting 
characteristic. It measures the amount of energy stored in the lattice deformation around
the electron. The polaron mass scales exponentially with $N_{\rm ph}$ so in addition the latter 
gives a rough estimate of the polaron mass. An estimator for $N_{\rm ph}$ can be derived by
noting that the last term in the Hamiltonian (\ref{eq:fifteen}) can be rewritten as
$\hbar \omega ( \hat{N}_{\rm ph} + \frac{1}{2} )$, where $\hat{N}_{\rm ph}$ is the operator
of the total number of phonons. The mean value of $\hat{N}_{\rm ph}$ is obtained by 
differentiating the free energy with respect to $\hbar \omega$. To make sure that no 
contribution is received from the interaction term, the derivative must be taken with the
combination $f^2/(M\omega)$ kept constant. Using $F = - \frac{1}{\beta} \ln Z$, the 
estimator is obtained as follows
\begin{equation}
N_{\rm ph} = - \frac{1}{\beta} 
\left. \frac{\partial F}{\partial (\hbar \omega)} \right\vert_{\frac{f^2}{M\omega}} = 
- \frac{1}{\beta} \left\langle \left. \frac{\partial A_{\rm pol}}{\partial (\hbar \omega)} 
\right\vert_{\frac{f^2}{M\omega}} \right\rangle_{\rm shift} .
\label{eq:thirtysix}
\end{equation} 
The subscript ``shift'' implies that the number of phonons corresponds to the ground state
of the polaron (${\bf K} = 0$). It is also possible to derive estimators for non-zero total momenta,
which is omitted here.

Finally, one can derive an estimator for the isotope exponent of the polaron effective mass 
$\alpha_{\mu}$. In the (bi)polaron mechanism of superconductivity, $\alpha_{\mu}$ is related
to the isotope effect on the critical temperature \cite{Alexandrov1992}. The mass isotope 
exponent is defined as $m^{\ast}_{\mu} = M^{\alpha_{\mu}}$, where $M$ is the ion mass. Since 
the present method calculates the inverse polaron mass, $\alpha_{\mu}$ is more conveniently
expressed via the inverse effective mass
\begin{equation}
\alpha_{\mu} = - \frac{M}{\frac{1}{m^{\ast}_{\mu}}} \frac{\partial}{\partial M}
\left( \frac{1}{m^{\ast}_{\mu}} \right) =
\frac{\omega}{2 \left( \frac{1}{m^{\ast}_{\mu}} \right) } \frac{\partial}{\partial \omega}
\left( \frac{1}{m^{\ast}_{\mu}} \right)   .
\label{eq:thirtyseven}
\end{equation} 
The last transformation follows from the scaling $M \propto \omega^{-2}$. The estimator for
$\alpha_{\mu}$ is derived directly from the estimator for the inverse effective mass (\ref{eq:six}).
A path's weight depends on the phonon frequency only via the polaron action $e^{A_{\rm pol}}$. 
(One should recall that the definition of a Monte Carlo average 
$\langle ( \Delta r_{\mu} )^2 \rangle$ contains $A_{\rm pol}$ in the numerator and denominator.)
Upon differentiation one obtains
\begin{equation}
\alpha_{\mu} = \frac{\omega}{2 \langle ( \Delta r_{\mu} )^2 \rangle}
\left[ \left\langle ( \Delta r_{\mu} )^2 \left. \frac{\partial A_{\rm pol}}{\partial \omega} 
\right\vert _{M \omega^2} \right\rangle  - \langle ( \Delta r_{\mu} )^2 \rangle
\left\langle \left. \frac{\partial A_{\rm pol}}{\partial \omega} 
\right\vert _{M \omega^2} \right\rangle \right] .
\label{eq:thirtyeight}
\end{equation}
In taking the frequency derivative, one should keep the combination $M\omega^2$ constant, as
it is independent of the ion mass. As with the phonon action and potential energy, the estimators
for the number of phonons and isotope exponent can be split into a double sum over path's 
segments. The expressions for the respective summands can be found in \cite{Spenser2005}.

\section{ \label{sec:four}
Polaron properties
}

The focus of this section is going to be the polaron properties that have been obtained 
with the continuous-time path-integral Quantum Monte Carlo method. For each topic, additional 
details of the algorithm will be provided, if necessary, on top of the general description 
presented above.

\begin{figure}
\centering
\includegraphics[width=11.7cm]{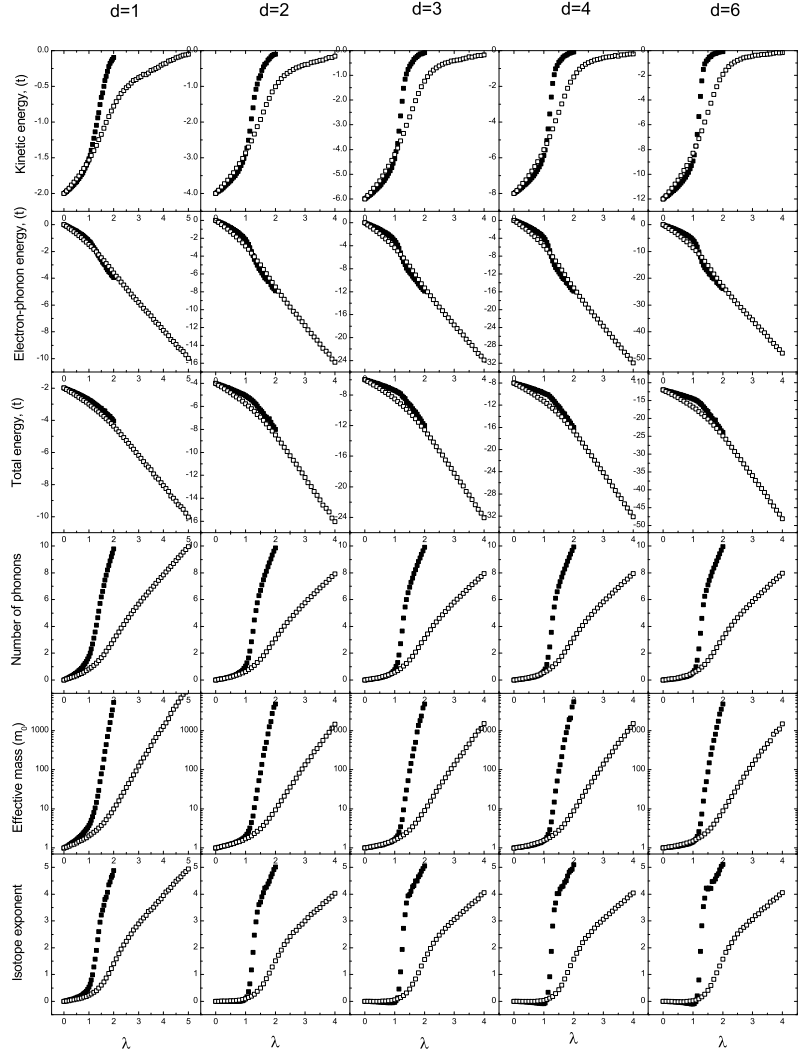}
\caption{
Properties of the Holstein polaron on $d=1, 2, 3, 4, 6$ hypercubic lattices. 
The filled and open symbols show data for the phonon frequencies $\hbar\omega/t = 0.4$ and 1.0 
($d=1$), 0.8 and 2.0 ($d=2$), 1.2 and 3.0 ($d=3$), 1.6 and 4.0 ($d=4$), and 2.4 and 6.0 ($d=6$), 
respectively. The coupling constant is defined as $\lambda = E_p/(zt) = \kappa^2/(2M\omega^2 zt)$,
where $z=2d$ is the number of nearest neighbours. The effective mass is measured in units of
$m_0 = \hbar^2/(2ta^2)$. Note that because the phonon frequency increases with $d$, the curves in 
different dimensions have similar shapes except for small differences around the $\lambda$ of 
polaron formation.
}
\label{fig:Holstein}       
\end{figure}

\subsection{ \label{sec:fourone}
Long-range interaction produces light polarons
} 

The Holstein model \cite{Holstein1959} describes polarons in systems with well-screened 
short-range electron-phonon interaction, such as molecular crystals. It is in the short list of 
``main'' polaron models. Due to its simplicity, the model has been extremely popular with the 
numerical community. Almost any new numerical method was first tried on the Holstein model, for 
which a lot of accurate results are available. Also, for a long time the Holstein model was 
regarded as a ``generic'' model for all narrow-band systems where the e-ph and lattice effects 
were equally important. The locality of the interaction was considered simply a matter of 
convenience, and the Holstein model itself was believed to contain all the qualitative features 
of lattice polarons. This turns out not to be the case. The main reason lies in the very sharp, 
exponential, dependence on the model parameters of the main polaron characteristic, the 
effective mass. Some properties of the Holstein polaron are presented in Fig.~\ref{fig:Holstein}. 
Any approximation or model simplification that have little effect on the polaron energy can have 
dramatic consequences for the mass, leading to false qualitative conclusions. In particular, the 
Holstein model is {\em not} adequate for complex oxides of transition metals such as the cuprates 
or manganites. In those materials the bare bands are narrow, $\sim 1 - 2$ eV, so the lattice 
effects are important. At the same time, a low density of free carriers cannot fully screen the 
Coulomb interaction between electrons and distant anions and cations, leading to a long-range 
e-ph interaction. 

\begin{figure}[b]
\centering
\includegraphics[width=11.7cm]{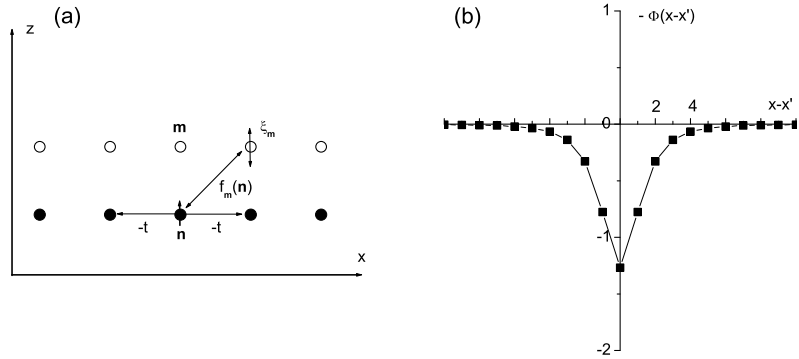}
\vspace{-0.5cm}
\caption{
(a) One-dimensional model with a long-range electron-phonon interaction. The electron 
moves along the bottom chain of sites shown by full circles (denoted by ${\bf n}$). 
The vibrating ions shown by open circles oscillate along the $z$-axis. The interaction 
is characterized by the $z$-projection of the Coulombic force (\ref{eq:thirtynine}). 
(b) Spatial profile of function $\Phi$, (\ref{eq:thirtythree}), that characterizes 
the retarded interaction of different parts of the polaron path. The parameters are 
$\kappa = 1$ and $d = a$. {\em Negative} $\Phi$ is shown in order to draw analogy with 
a potential well. This shape of $\Phi$ should be contrasted with that of the Holstein 
model $\Phi_{\rm Hol} = \kappa^2 \delta_{x x'}$. A smooth $\Phi$ cannot localize the 
path as well as a sharp one, which results in a smaller effective mass in the case of 
a long-range model.
}
\label{fig:three}       
\end{figure}

A model of this kind was considered a long time ago by Eagles \cite{Eagles}. 
(The corresponding polaron was called there a ``nearly-small polaron''.) More recently, a 
long-range model with linearly polarized phonons was put forward in relation to high-temperature 
superconductors in \cite{longrange}. The model is depicted in Fig.~\ref{fig:three}(a). 
An electron hops within a rigid chain (in the one-dimensional case) or plane (in the 
two-dimensional case) and interacts with ions that are positioned above the electron 
chain (or plane) and vibrate perpendicular to the chain (plane). Such a polarization was chosen 
to represent the $c$-polarized phonon modes that feature prominently in the cuprates 
\cite{Timusk1995}. The force function was taken to be the Coulomb force projected on the 
$z$-axis
\begin{equation}
f_{\bf m}({\bf n}) = \kappa a^2 \frac{({\bf n}-{\bf m})_z}{\vert {\bf n}-{\bf m} \vert^3} = 
\frac{\kappa a^2 d}{[({\bf n} - {\bf n}_{\bf m})^2 + d^2]^{3/2}} \: .
\label{eq:thirtynine}
\end{equation}
Here $a$ is the lattice constant, $d$ is the distance between the vibrating ions and the electron
plane, and ${\bf n}_{\bf m}$ is the vector of the electron Wannier orbital in the same unit cell
as ion ${\bf m}$. The interaction strength is characterized by the parameter $\kappa$.  
Alternatively, one can use a dimensionless parameter 
\begin{equation}
\lambda = \frac{E_p}{zt} = \frac{1}{2 M \omega^2 z t} \sum_{\bf m} f^2_{\bf m}(0) \: ,
\label{eq:lambda}
\end{equation}
which is a ratio of the polaron energy in the atomic limit $E_p$ (at $t = 0$) to the kinetic 
energy of a free electron $zt$ (at $\kappa = 0$). Here $z$ is the number of nearest lattice sites. 
The confinement of the polaron path is determined by the function $\Phi$, 
(\ref{eq:thirtythree}), whose profile is shown in Fig.~\ref{fig:three}(b). (It should be 
compared with the Holstein confining function $\kappa^2 \delta_{\bf nn'}$.)  Due to loose confinement, 
the polaron path diffused farther than in the Holstein case, which is equivalent to a much lighter 
mass. Such a polaron was named the ``small Fr\"ohlich polaron'' (SFP) in \cite{longrange}. 

\begin{figure}
\centering
\vspace{-0.5cm}
\includegraphics[width=11.7cm]{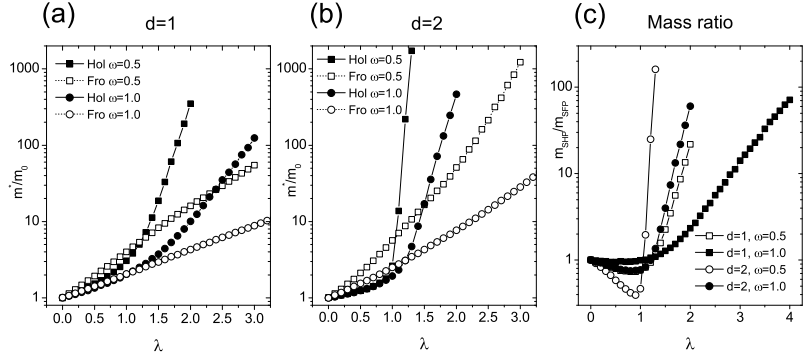}
\vspace{-0.5cm}
\caption{
Effective masses of small Holstein and Fr\"ohlich polarons in units of $m_0 = \hbar^2/(2ta^2)$. 
(a) $d = 1$. (b) $d = 2$. 
(c) The mass ratio of the small Holstein and Fr\"ohlich polarons for several model parameters. 
The ratio scales exponentially with the coupling, and could exceed 100. 
}
\label{fig:four}       
\end{figure}

The masses of SFP and SHP calculated with PIQMC are compared in Fig.~\ref{fig:four}. SFP is 
slightly heavier at small $\lambda$ but much lighter at large $\lambda$. The ratio of the two 
masses is a non-monotonic function. This observation was later confirmed by exact diagonalization 
\cite{Fehske2000} and variational \cite{Bonca2001,Perroni2004} methods. The most interesting 
property of SFP is exponential reduction of mass relative to SHP, see Fig.~\ref{fig:four}(c). 
This effect is independent of the dimensionality because it originates solely from the 
long-ranginess of e-ph interaction. The mass reduction can be very large. For example, in two 
dimensions at $\hbar\omega/ t = 0.5$ and $\lambda = 1.2$, $m_{\rm SHP} \approx 220$ while 
$m_{\rm SFP}$ is only $\approx 9$. In \cite{Spenser2005}, an additional screening factor was 
added to the force function (\ref{eq:thirtynine}). All polaron properties smoothly interpolated 
between those of SHP and SFP as the screening radius was changed from zero to infinity.             

The physical reason for the small mass of SFP is easy to understand. The mass is determined by the
overlap of the ionic wave functions before and after the electron hop. In the case of short-range
e-ph interaction, the lattice deformation must relax all the way back to the equilibrium state 
after the electron leaves the site. Likewise, the lattice deformation at the new electron's location
must form anew from the equilibrium state. As a result, the overlap integral is exponentially
small. In the case of a long-range interaction, the deformation at the new location is partially
pre-existent. The ions must move by a smaller distance, which results in a much larger overlap of
the wave functions. That also produces an exponentially large mass, but with a smaller exponent.
(The fact that the mass is still exponential in coupling justifies the term ``small'' in the 
polaron's name.) This effect can also be understood within the Lang-Firsov (LF) approximation 
for the polaron mass \cite{Tjablikov1952,Firsov1962} that becomes exact in the limit of 
infinitely fast ions, $\hbar \omega \gg t$. According to LF, the polaron mass is given by 
$m^{\ast} = m_0 \, e^{\gamma (E_p/\hbar \omega)}$, where $E_p$ is the polaron shift from 
(\ref{eq:lambda}). The dimensionless parameter
\begin{equation}
\gamma = 1 - 
\frac{\sum_{\bf m} f_{\bf m}(0) f_{\bf m}({\bf a})}{\sum_{\bf m} f^2_{\bf m}(0) } \: ,
\label{eq:gamma}
\end{equation}
depends on the shape of the e-ph interaction. For the force function (\ref{eq:thirtynine}) at 
$d = a$, $\gamma = 0.387$ for the linear chain, $0.334$ for the square lattice and $0.320$ for 
the triangular lattice. Note also that a long-range e-ph interaction smoothens the polaron 
crossover, cf. Fig.~\ref{fig:four}, which makes the single exponential form be more applicable 
for the entire coupling range \cite{longrange}. 

These results demonstrate that the Holstein model is an {\em extreme} model that predicts the 
largest mass for the given polaron energy. As such, the Holstein model is not adequate for 
ionic systems with poor screening where the e-ph interaction is long range.

\subsection{  \label{sec:fourtwo}
Enhancement of anisotropy by electron-phonon interaction
}

\begin{figure}
\centering
\includegraphics[width=11.7cm]{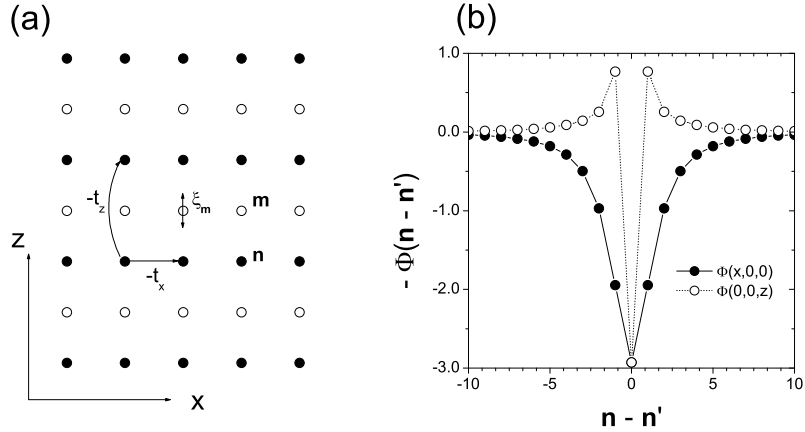}
\vspace{-0.5cm}
\caption{
(a) A three-dimensional polaron model with anisotropic e-ph interaction. The electron moves
within and between the planes of filled circles with hopping integrals $-t_x$ and $-t_z$,
respectively. It interacts with vibrating ions shown by open circles. The vibrations are 
polarized along the $z$-axis. In the case of a long-range interaction the polaron is 
Fr\"ohlich-like in the $xy$ plane and Holstein-like along the $z$ direction.  
(b) The (negative) polaron path confinement function $- \kappa^{-2} \Phi (x,y,z)$. 
An additional factor $e^{-{|{\bf n} - {\bf m}|}/R}$ with $R = 10$ lattice constants in the
$x$ direction, has been added to the force function (\ref{eq:thirtynine}) to help the lattice 
sum to converge. The confining profile along $z$ is much steeper than along $x$. 
}
\label{fig:anisotmodel}       
\end{figure}

Two different behaviours of the lattice polaron can lead to an interesting situation 
when the same particle is Holstein-like and Fr\"ohlich-like along two different lattice directions. 
A concrete model that features such properties was put forward in \cite{anisotropy}. The model is 
an extension of the force function (\ref{eq:thirtynine}) to three dimensions, see 
Fig.~\ref{fig:anisotmodel}(a). Clearly, the e-ph interaction is anisotropic, which translates to an 
even higher anisotropy of polaron spectrum. Upon hops along the $z$ direction, the electron must 
reverse the sign of deformation. Thus the ion overlap integrals are even smaller than in the 
Holstein case. In contrast, the movement in the $xy$ plane is Fr\"ohlich-like, as described in the 
previous section. The deformation is partially prepared, the overlap integrals are large, and the 
effective mass is small. This reasoning is corroborated by the spatial profiles of the path
confinement function $\Phi$, see Fig.~\ref{fig:anisotmodel}(b). When the path diffuses along the
$x$ direction, $\Phi$ is smooth which translates into easy ``escape''. Along the $z$ direction,
$\Phi$ is much steeper. In fact it even changes sign upon migration to the nearest plane, which
reflects the sign reversal of the lattice deformation after the electron hop. Since $\Phi$ affects 
the path's weight exponentially, one expects exponentially large difference between the polaron
masses in $z$ and $x$ direction:
$m^{\ast}_z/m^{\ast}_x \propto e^{(\gamma_z - \gamma_x) \lambda / \bar{\omega}}$, where 
$\lambda = 2.93 \kappa^2/(12 M \omega^2 t_x)$ and $\bar{\omega} = \hbar\omega/t_x$ is the 
dimensionless phonon frequency. As a result, the mass anisotropy grows exponentially with
$\lambda$ and can reach very big values. In addition, the mass anisotropy is a sharp
function of the phonon frequency, which should lead, for example, to an isotope effect 
on anisotropy.   

\vspace{-0.5cm}

\begin{figure}
\centering
\includegraphics[width=8.0cm]{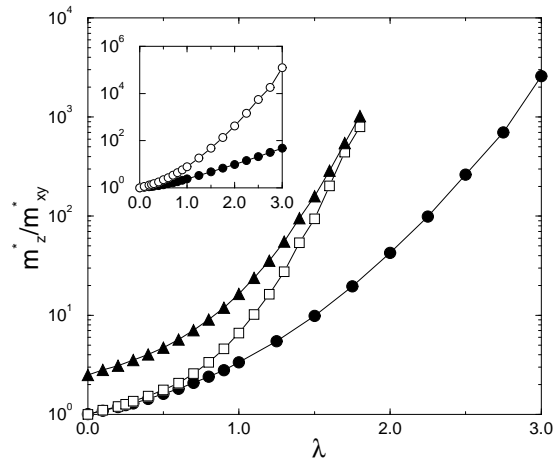}
\caption{
Mass anisotropy of the three-dimensional model (\ref{eq:thirtynine}).   
Circles:   $\hbar\omega = 1.0 \, t_x$, $t_z = 0.25 \, t_x$;
squares:   $\hbar\omega = 0.5 \, t_x$, $t_z = 0.25 \, t_x$;
triangles: $\hbar\omega = 0.5 \, t_x$, $t_z = 0.1  \, t_x$.
Inset: masses $m^{\ast}_{xy}$ (filled circles) and $m^{\ast}_z$ for 
$\hbar\omega = 1.0 \, t_x$, $t_z = 0.25 \, t_x$. From \cite{anisotropy}. 
}
\label{fig:anisotresults}       
\end{figure}

The results of PIQMC analysis of the model are summarized in Fig.~\ref{fig:anisotresults}
\cite{anisotropy}. The inset shows a typical behaviour of the polaron masses at 
$\hbar\omega = 1.0 \, t_x$ and $t_z = 0.25 \, t_x$. (This choice of hoppings ensures the 
isotropy of the bare spectrum, since the lattice constant in $z$ direction is assumed to be 
twice the one in $x$ direction.) As expected, $m^{\ast}_x$ grows exponentially with coupling. 
Interestingly, the $z$ mass grows {\em super}-exponentially with a large quadratic component: 
$m^{\ast}_z \approx m_{x0} \, e^{1.26 \lambda + 0.88 \lambda^2}$. It is very possible though 
that this is still transient regime, and $m^{\ast}_z$ approaches pure exponential growth at 
still larger $\lambda$ (at which $m_z$ becomes so large it is difficult to calculate).
The mass anisotropy for several sets of model parameters is shown in the main panel of 
Fig.~\ref{fig:anisotresults}. Due to the super-exponential growth of $m^{\ast}_z$, the
anisotropy is also super-exponential, e.g. 
$m^{\ast}_z/m^{\ast}_{xy} \approx e^{0.42 \lambda + 0.71 \lambda^2}$ for 
$\hbar\omega = 1.0 \, t_x$ (circles). At a smaller frequency $\hbar\omega = 0.5 \, t_x$ 
(squares) the anisotropy is exponentially larger, as expected from the reasoning given above. 
This implies the existence of an {\em isotope effect on the mass anisotropy}. The third model
parameter, the bare hopping anisotropy, turns out to have little effect on the anisotropy of
the polaron spectrum. The mass anisotropy for $t_z = 0.1 \, t_x$ and the same phonon
frequency $\hbar\omega = 0.5 \, t_x$ is shown by triangles in Fig.~\ref{fig:anisotresults}.
While being $2.0 - 2.5$ times larger at small $\lambda$, the anisotropy approaches that
of the $t_z = 0.25 \, t_x$ case at large $\lambda$. It shows that it is primarily the e-ph 
interaction that governs the polaron anisotropy. This conclusion is further supported by
studies of the Holstein model with anisotropic bare hopping \cite{anisotropy} where {\em no} 
enhancement of the polaron anisotropy was observed.

In relation to the effect described, it is interesting to note a well-do\-cu\-men\-ted discrepancy 
between the theoretical and experimental anisotropy of the cup\-rates \cite{Cooper1994}. According 
to band structure calculations, the anisotropy of resistivity of LSCO and YBCO compounds should 
be about $10 - 30$ \cite{Allen1987,Pickett1989}. At the same time, the experimental anisotropy of 
resistivity is between $10^2$ and $10^3$ depending on the level of doping. The anisotropy of
bismuthates is even higher, $(5 - 80)\cdot 10^4$ \cite{Cooper1994}, which is difficult to
explain with the conventional Bloch-Boltzmann theory. According to the present results, 
anisotropic interaction with $z$-polarized phonons is a sufficient condition for a very large $z$ 
effective mass. Of course, the anisotropy of mass and resistivity are two different things, but it 
would be fair to assume that the former is at least partially responsible for the latter 
(see, e.g., \cite{Cooper1994} page 85). This idea still awaits proper development. Alternative 
explanations of the anomalous $z$-transport exist as well \cite{AKM1996}.

\subsection{  \label{sec:fourthree}
Spectrum flattening and polaron density of states
}

\begin{figure}[t]
\centering
\includegraphics[width=11.7cm]{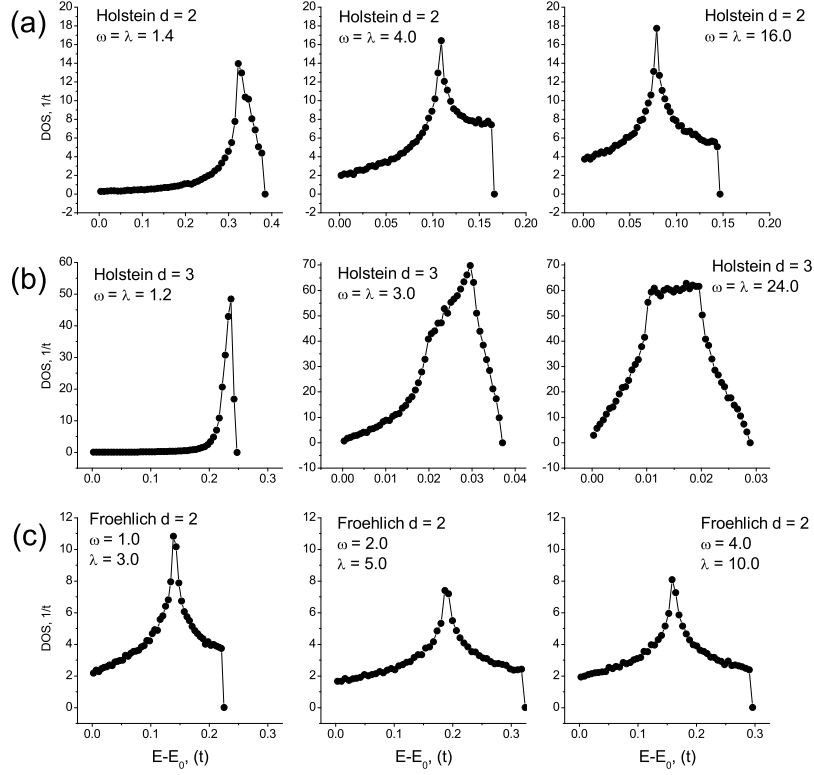}
\vspace{-0.5cm}
\caption{
(a) and (b) The evolution of the density of states of the Holstein polaron with phonon 
frequency $\omega$ in $d = 2$ and $3$, respectively. Each graph was obtained by calculating 
the polaron spectrum at 100,000 ${\bf K}$ points randomly distributed in the Brillouin zone, 
and histogramming the results between 50 energy bins. Each spectrum point was calculated 
by averaging 250,000 values of $\cos{{\bf K} \Delta {\bf r}}$ taken every 10th path update. 
Every 5000 measurements the path was reset and then equilibrated for 1000 updates.   
(c) The same for the small Fr\"ohlich polaron in $d = 2$.
}
\label{fig:DOS}       
\end{figure}

A great power of the PIQMC method is the ability to calculate an entire polaron spectrum in any
dimensionality in a single run, as was explained in Section~\ref{sec:threethree}. This opens up
an exciting possibility to calculate the polaron density of states (DOS) 
$\rho(E) = N^{-1} \sum_{\bf K} \delta(E - E_{\bf K} + E_0)$ by discretizing the energy interval
and histogramming $E_{\bf K}$ values at the end of simulations. The polaron spectrum in the 
adiabatic limit ($t \gg \hbar \omega$) possesses an interesting property of flattening at large 
lattice momenta, which has been well documented in the literature 
\cite{Wellein1996,Wellein1997,Romero1998,Stephan1996}. In the weak-coupling limit, the flattening 
can be readily understood as hybridization between the bare electron spectrum and a
momentum-independent phonon mode. The resulting polaron dispersion is cosine-like at small 
${\bf K}$ and flat at large ${\bf K}$. As larger couplings this picture becomes less and less 
intuitive but the flattening remains as a matter of fact \cite{Levinson1973}. It is noteworthy 
that in large spatial dimensions the phase space is dominated by states with large momenta. If 
all those states have close energies, they should form a peak in the DOS close to the {\em top} 
of the polaron band. Exact PIQMC calculations have confirmed that this is indeed the case 
\cite{Kornilovitch1999}. The evolution of the DOS of the isotropic Holstein model with phonon 
frequency $\omega$ in two and three dimensions is shown in Fig.~\ref{fig:DOS}(a) and (b). 
In all the cases presented the polaron is fully developed, with the bandwidth much smaller than 
$\omega$.

At small $\omega$, DOS develops a massive peak at the top of the band. The peak is more 
pronounced in $d=3$ than in $d=2$. The van Hove singularities are absorbed in the peak and as 
such cannot be seen. With increasing $\omega$, the polaron spectrum approached the cosine-like 
shape in full accordance with the Lang-Firsov non-adiabatic formula. The respective DOS gradually 
assume the familiar shapes of tight-binding zones with renormalized hopping integrals. The van 
Hove singularities are clearly visible. These results have an interesting corollary for the
Holstein model. At small-to-moderate $\omega$ in two and three dimensions, the bottom half of 
the polaron band contains a tiny minority of the total number of states (measured in low
percentage points). In most real systems those states will be localized and hence irrelevant. 
All the system's responses will be dominated by the states in the peak. The physical properties 
of these physically relevant states are going to be quite different from those of the ground 
state that are usually analyzed theoretically.   

It is interesting to look at the DOS of the long-range model (\ref{eq:thirtynine}). The two 
dimensional DOS is shown in Fig.~\ref{fig:DOS}(c). It is much closer to the 
tight-binding shape than the Holstein DOS at the same parameters.  The polaron spectrum and 
density of states is another manifestation of the extremity of the Holstein model.
Long-range e-ph interactions remove those peculiarities and make the polaron bands more 
``normal''.  

Some densities of states of the Holstein model with anisotropic hopping were presented in
\cite{Kornilovitch1999}.

\subsection{  \label{sec:fourfour}
Isotope exponents
}

Isotope substitution is a powerful tool in determining if a particular phenomenon or feature 
is of phonon origin. Some polaron properties, such as mass, are sharp functions of the lattice 
parameters, therefore isotope effects in (bi)polarons are strongly enhanced. In particular, 
anything that depends on the (bi)polaron mass should exhibit a large isotope exponent 
\cite{Alexandrov1992}. A large isotope effect was observed, for example, on the magnetic 
penetration depth in cuprates \cite{Zhao1997,Khasanov2004,Zhao2006}, which was interpreted 
as evidence for bipolaronic carriers in the superconducting state.   

\begin{figure}[b]
\centering
\vspace{-0.5cm}
\includegraphics[width=11.7cm]{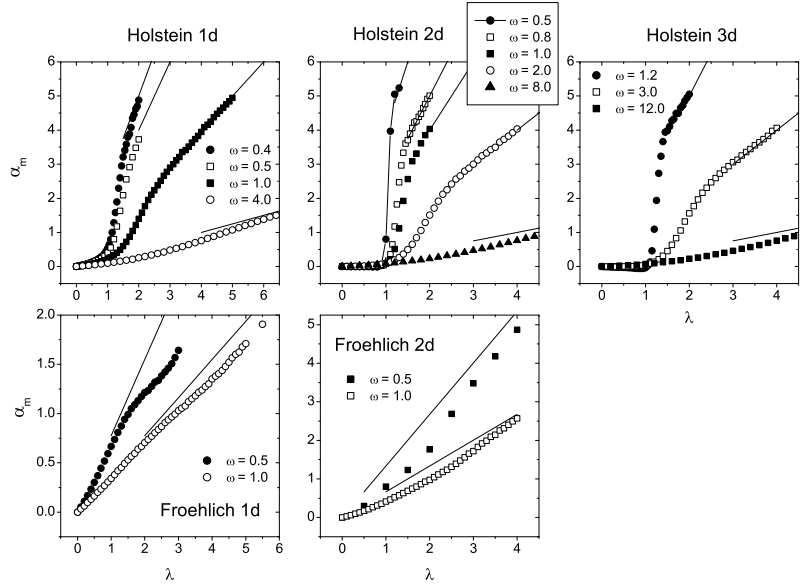}
\vspace{-0.5cm}
\caption{
Mass isotope exponents of the small Holstein and small Fr\"ohlich polarons in different
dimensions and for different phonon frequencies. The thin solid lines indicate the strong-coupling
limit $\alpha = \frac{\gamma}{2} \frac{\lambda zt}{\hbar\omega}$. Note that $\alpha$ of the
Holstein polaron in $d = 2$ and $3$ is {\em negative} at small $\omega$.
}
\label{fig:IsotopeMass}       
\end{figure}

As was explained in Section~\ref{sec:threethree}, the PIQMC method enables appro\-xi\-ma\-tion-free 
calculation of the isotope exponent on the (bi)polaron effective mass for a large class of 
e-ph models. A good feel for $\alpha$ in the strong-coupling limit can be obtained from the 
anti-adiabatic expression for the polaron mass, $m^{\ast} = m_0 \, e^{\gamma ( E_p/\hbar\omega)}$. 
Since the polaron shift $E_p$ is ion-mass-independent, 
$\alpha = \frac{\gamma}{2} \frac{E_p}{\hbar\omega}$. Thus $\alpha$ is proportional to the 
coupling constant $\lambda$. In the weak coupling regime, $\alpha$ can be computed 
perturbatively. For example, for the one-dimensional Holstein model, the second order yields
\begin{equation}
\alpha^{(2)}_{\rm 1d \: Holstein} = 
\lambda \frac{\bar{\omega}^2 (\bar{\omega}^2 + 2 \bar{\omega} + 4)}
             {(\bar{\omega}^2 + 4 \bar{\omega})^{\frac{5}{2}}} \: ,
\label{eq:forty}
\end{equation}
where $\bar\omega = \hbar\omega/t$. (The second-order coefficients of other polaron
properties of the one-dimensional Holstein model can be found, e.g., in \cite{Spenser2005}.
The second order coefficients for higher-dimensional lattices were given in \cite{Hague2005}.)
The isotope exponent is again proportional to $\lambda$ but with a different coefficient. 
The two linear dependencies should be smoothly connected over the polaron crossover. 

The mass isotope exponents for the small Holstein polaron are shown in the top row of 
Fig.~\ref{fig:IsotopeMass}. Before the polaron transition $\alpha$ is small reflecting the
non-polaronic character of the carrier. Note that in $d=2$ and $3$, $\alpha$ is {\em negative} 
at small frequencies \cite{Hague2005}. After the polaron crossover, which always begins 
at $\lambda \sim 1$ but ends at $\lambda$ that increases with $\omega$, the isotope exponent
quickly reaches the strong-coupling asymptotic behaviour with $\gamma = 1$. Notice that the
beginning and the end of the transition are clearly identifiable on the plots. Thus the 
mass isotope exponent can be useful in analyzing the Holstein polaron transition.

The case of the small Fr\"ohlich polaron is somewhat different, see the bottom row of 
Fig.~\ref{fig:IsotopeMass}. At small coupling, the exponent grows linearly with a $\gamma$ 
close to the strong coupling limit, but then deviates to smaller values. The exponent returns 
back to the strong-coupling limit at much larger $\lambda$, whose value decreases with 
increasing $\omega$. This final approach happens when the entire path is mostly confined to 
one lattice site, and only rarely deviates to the first nearest neighbour. This interesting 
behaviour is not yet fully understood.

\begin{figure}
\centering
\includegraphics[width=11.7cm]{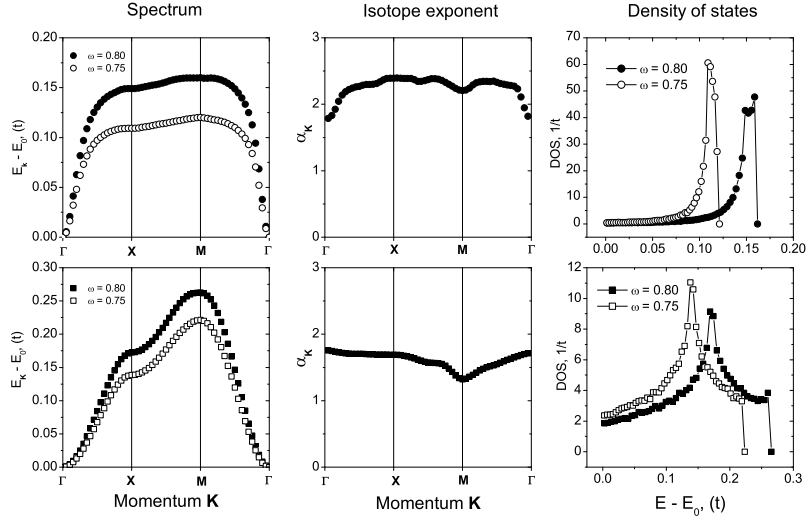}
\caption{
Isotope effect on polaron spectrum and density of states. 
{\em Top row}\/: the $2d$ Holstein polaron at $\lambda = 1.2$. Left: polaron spectrum at 
$\bar{\omega} = 0.80$ and $0.75$. Middle: the isotope exponent for each ${\bf K}$-point. 
Right: the density of states for the two frequencies.  
{\em Bottom row}\/: the same for the small $2d$ Fr\"ohlich polaron at $\lambda = 2.4$.
}
\label{fig:IsotopeSpectrum}       
\end{figure}

As was shown in the previous section, in the adiabatic regime the properties of the polaron 
mass do not fully represent those of the vast majority of polaron states. Additional insight 
can be gained from an isotope effect on the polaron bandwidth or even on individual spectrum 
points \cite{isotope}. The isotope effect on polaron spectrum and density of states in $d=2$ 
is illustrated in Fig~\ref{fig:IsotopeSpectrum}. The ratio of the two phonon frequencies, 
$\hbar\omega = 0.80 \, t$ and $0.75 \, t$, has been chosen to roughly correspond to the 
substitution of $^{16}$O for $^{18}$O in complex oxides. One can see that the polaron band
shrinks significantly, by 20-30\%, for both polaron types. The middle panels show the isotope 
exponents on spectrum points calculated as
\begin{equation}
\alpha_{\bf K} = \frac{1}{2} \frac{\langle \omega \rangle}{\langle E_{\bf K} \rangle} 
\frac{\Delta E_{\bf K}}{\Delta \omega} \: ,
\label{eq:fortyoneone}
\end{equation}
where the angular brackets denote the mean value of either the two frequencies or of the two 
energy values. An interesting observation is that $\alpha_{\bf K}$ of the Fr\"ohlich polaron
is roughly independent of ${\bf K}$ ($\pm 10\%$). In the Holstein case $\alpha_{\bf K}$ dips
in the vicinity of the ${\rm \Gamma}$ point. (Small fluctuations at intermediate momenta are
due to statistical errors in averaging $\langle \cos {\bf K} \Delta {\bf r} \rangle$.)

\subsection{  \label{sec:fourfive}
Jahn-Teller polaron
}

Density-displacement is only one possible type of e-ph interaction. Examples of other types 
are the deformation potential and Su-Schrieffer-Heeger (SSH) interaction. In the former, the 
electron density interacts with a gradient of lattice displacement whereas in the SSH case the 
lattice deformation is coupled to electron's kinetic energy. The Jahn-Teller (JT) interaction 
is one of the most complex types because it usually involves a multidimensional electron basis 
and a multidimensional representation of the deformation group. The JT interaction is active in 
some molecules and crystals of high point symmetry. The JT effect was also a guiding principle 
in the search for high-temperature superconductivity \cite{Bednorz1988}. More recently, a JT 
theory of the cuprates was developed in \cite{Mihailovic2001,Kabanov2002,Mertelj2005}. 
There exist different flavors of the JT interaction. The simplest one is the $E \otimes \, e$ 
interaction \cite{Kanamori1960} that describes a short-range coupling between twice-degenerate 
$e_g$ electronic levels $(c_1, c_2)$ and a local double-degenerate vibron mode $(\zeta,\eta)$. 
The Hamiltonian reads
\begin{eqnarray}
H_{\rm JT} = & - & t \sum_{\langle {\bf n n'} \rangle } \left( 
c^{\dagger}_{{\bf n'}1} c_{{\bf n}1} + c^{\dagger}_{{\bf n'}2} c_{{\bf n}2} \right) 
\nonumber \\
& - & \kappa \sum_{\bf n} \left[ 
  \left( c^{\dagger}_{{\bf n}1} c_{{\bf n}2} + c^{\dagger}_{{\bf n}2} c_{{\bf n}1} \right) \eta_{\bf n} 
+ \left( c^{\dagger}_{{\bf n}1} c_{{\bf n}1} - c^{\dagger}_{{\bf n}2} c_{{\bf n}2} \right) \zeta_{\bf n}
\right] \nonumber \\
& + & \sum_{\bf n} \left[ - \frac{\hbar^2}{2M} \left( \frac{\partial^2}{\partial \zeta^2_{\bf n}} + 
\frac{\partial^2}{\partial \eta^2_{\bf n}} \right) + 
\frac{M \omega^2}{2} \left( \zeta^2_{\bf n} + \eta^2_{\bf n} \right) \right] \: .
\label{eq:fortyone}
\end{eqnarray} 
The symmetry of the interaction ensures the same coupling parameter $\kappa$ for the two
phonons. The kinetic energy is chosen to connect like electron orbitals of the nearest 
neighbours. This choice is somewhat arbitrary (makes it more ``Holstein-like'') and not 
dictated by symmetry. Because the ionic coordinates of different cells are not coupled, 
the model describes a collection of separate clusters that are linked only by electron 
hopping. To relate the Hamiltonian to more realistic situations, phonon dispersion must 
be added \cite{Mihailovic2001,Allen1999}. 

An important property of the $E \otimes \, e$ interaction is the absence of an exact 
analytical solution in the atomic limit $t = 0$. In the density-displacement case
the dynamics of each $\zeta$ is described by a one-dimensional differential equation under 
a constant force. That has a shifted oscillator solution, which serves as a convenient 
starting point for various strong-coupling expansions. Here, in contrast, the atomic 
limit is described by two coupled {\em partial} differential equations for the electron 
doublet $\psi_{1,2}(\zeta,\eta)$. Although it is possible to separate variables and reduce 
the system to two ordinary second-order equations, they seem to be too complex to admit 
an analytical solution. At large couplings, however, the elastic energy assumes the
Mexican hat shape and the phonon dynamics separates into radial oscillatory motion and 
azimythal rotary motion. This results in an additional pre-exponential factor 
$\propto \kappa$ in the ion overlap integral, leading to the effective mass 
$m^{\ast}_{\rm JT} = m_0 \sqrt{\frac{2}{\pi g}} \, e^{g^2}$, where 
$g^2 = \frac{\kappa^2}{2M\hbar\omega^3}$ \cite{Takada2000}.

\begin{figure}
\centering
\includegraphics[width=11.7cm]{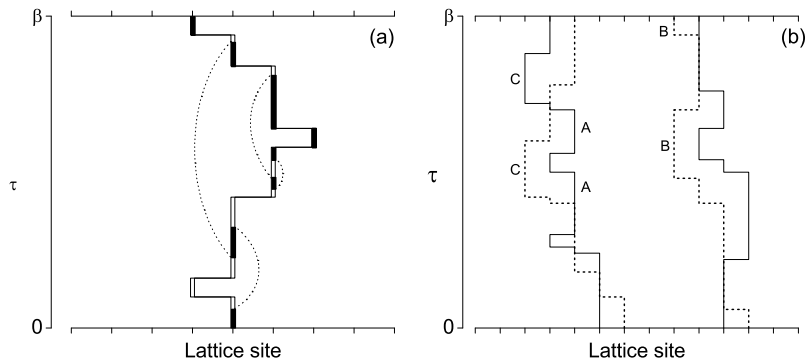}
\caption{Extensions of the basic polaron PIQMC method. 
(a) In the Jahn-Teller polaron, the path carries an internal orbital index shown as black or
white sections of the path. The $\eta$-phonon parings are represented by dashed lines. Each 
vertex changes the index.
(b) In the bipolaron case, the system is represented by two imaginary-time paths. A direct
configuration is shown on the left and an exchange one on the right. The interaction of 
segments marked A or B contribute to the energy of individual polarons. The segments that 
belong to different paths, such as those marked C, contribute to the interaction between
the polarons.}
\label{fig:nine}       
\end{figure}

A path integral approach to Hamiltonian (\ref{eq:fortyone}) was developed in 
\cite{Kornilovitch2000}. Because there are two electron orbitals, the electron path must 
be assigned an additional orbital index (or \textit{colour}) $a = 1,2$. Colour 1 (or 2) of 
a given path segment means that it resides in the first (second) atomic orbital of the 
electron doublet. The short-term density matrix is 
\begin{eqnarray}
\rho (\Delta \tau) & = & 
\langle {\bf r'} a'; \{ \zeta'_{\bf n} \} \{ \eta'_{\bf n} \} \vert e^{- \Delta \tau H} \vert 
{\bf r} a; \{ \zeta_{\bf n} \} \{ \eta_{\bf n} \} \rangle  \nonumber \\
& = & \left[ \delta_{\bf rr'} \delta_{aa'} + 
           ( \kappa \Delta \tau ) \delta_{\bf rr'} \delta_{a{\bar a}'} \eta_{\bf n} + 
( t \Delta \tau ) \delta_{a a'} \sum_{\bf l} \delta_{{\bf r'}, {\bf r}+ {\bf l}} \right] 
e^{A_{\rm ph}(\Delta \tau)} ,
\label{eq:fortytwo}
\end{eqnarray} 
\begin{eqnarray}
& & \hspace{-0.5cm}  A_{\rm ph}(\Delta \tau) =  
( \kappa \Delta \tau ) (\delta_{a1} - \delta_{a2} ) \zeta_{\bf n} \nonumber \\
& & - \sum_{\bf n} \left\{ \frac{M}{2\hbar^2 (\Delta \tau)} 
[ ( \zeta_{\bf n} - \zeta'_{\bf n} )^2 + ( \eta_{\bf n} - \eta'_{\bf n} )^2 ] +
(\Delta \tau) \frac{M\omega^2}{2} ( \zeta^2_{\bf n} + \eta^2_{\bf n} ) \right\} .
\label{eq:fortythree}
\end{eqnarray} 
Here ${\bar a} = 1 (2)$ when $a = 2 (1)$, that is ${\bar a}$ is ``not'' $a$. This expression 
reveals a difference between the two phonons. Phonon $\zeta$ is coupled to electron density, 
like in the Holstein case. The difference from the Holstein is that the direction of the force 
changes to the opposite when the electron changes orbitals. In contrast, phonon $\eta$ is 
coupled to orbital changes themselves: the more often the electron changes orbitals, the more 
``active'' is $\eta$. (Discrete orbital changes are analogous to electron hops between discrete 
lattice sites, and as such are associated with ``kinetic orbital energy''. Phonon coupling to 
orbital changes is analogous to phonon coupling to electron hopping in the Su-Schrieffer-Heeger
interaction \cite{Su1979}.) Multiplication of a large number of the short-time propagators generates
multiple terms, each of which contains a finite number of lattice hops and orbital changes. On top 
of that, the exponents combine in a total phonon action $A_{\rm ph}(\beta)$ that comprises the 
free $\eta$ action $A_{\eta 0}$ and the action of $\zeta$ phonons under an external force 
[$A_{\zeta}$ is similar to (\ref{eq:twentynine})]. The next step is integration over the paths 
$\eta(\tau)$ and $\zeta(\tau)$. Integration over $\zeta(\tau)$ is standard and performed as 
described as for the density-displacement interaction. The result is a factor $e^{A_{\zeta}}$ in 
the path's statistical weight, where
\begin{equation}
A_{\zeta}[{\bf r}(\tau), a(\tau)] = \kappa^2 \int^{\beta}_0 \!\! \int^{\beta}_0 
{\rm d}\tau {\rm d} \tau' G(\tau - \tau') 
\left[ \delta_{a(\tau), a(\tau')} - \delta_{a(\tau), {\bar a}(\tau')} \right] \: ,
\label{eq:fortyfour}
\end{equation}
\begin{equation}
G(\tau - \tau') = \frac{\hbar}{2M\omega} \left[ 
e^{-\hbar \omega \vert \tau - \tau' \vert} \delta_{{\bf r}(\tau),{\bf r}(\tau')} +
e^{-\hbar \omega ( \beta - \vert \tau - \tau' \vert)} 
\delta_{{\bf r}(\tau),{\bf r}(\tau')+ {\rm sgn}(\tau - \tau') \Delta {\bf r}}
\right] \! .
\label{eq:fortyfive}
\end{equation}
The last expression is valid under the condition $e^{\beta \hbar \omega} \gg 1$. 
As expected, action $A_{\zeta}$ is an explicit functional of the spatial path ${\bf r}(\tau)$
and of the orbital path $a(\tau)$. Note that the $\zeta$ phonon favours like orbitals and 
disfavours orbital changes. Integration over $\eta(\tau)$ is trickier for it contains $\eta$ 
as pre-exponential factors. A typical term with $N_s$ orbital changes has the following form: 
\begin{equation}
(\kappa \Delta \tau)^n \eta_{{\bf r}(\tau_1)}(\tau_1) \eta_{{\bf r}(\tau_2)}(\tau_2)
\ldots \eta_{{\bf r}(\tau_{N_s})}(\tau_{N_s}) \, e^{A_{\eta 0}[\eta_{\bf n}(\tau)]} \: .
\label{eq:fortysix}
\end{equation}
Here, $\tau_s$ is the time of $s$th orbital change ($s = 1, \ldots , N_s$), ${\bf r}(\tau_s)$
is the electron position at this time, and $\eta_{{\bf r}(\tau_s)}(\tau_s)$ is the 
$\eta$-displacement at this site at this time. For odd $N_s$, path integration over 
$\eta_{\bf n}(\tau)$ produces zero by parity. For even $m$, the integration can be 
performed by introducing fictitious sources, calculating the generating functional and
differentiating it $m$ times. The result is a sum of all possible pairings of $\tau_s$. 
Within each term, each pair contributes the factor $\kappa^2 G(\tau_s - \tau_{s'})$, with
$G$ given by (\ref{eq:fortyfive}). Since $G$ is positive, the statistical weight
increases with the number of orbital flips. Thus, the $\eta$-phonon {\em favours}
orbital flips, in complete contrast with the $\zeta$-phonon. Since both phonons are
governed by the same Green's function (which should be no surprise since the two
phonons are related by symmetry) one should expect a dynamical equilibrium between
the two phonons and some finite mean value of orbital flips for given coupling.
The shift partition function is    
\begin{equation}
Z_{\Delta {\bf r}} = Z_{\rm ph} \sum^{\infty}_{N_k = 0,1, \ldots} 
\sum^{\infty}_{N_s = 0,2, \ldots}  \int^{\beta}_{0} \cdots \int^{\beta}_{0} 
({\rm d}\tau)^{N_k} ({\rm d}\tau)^{N_s} W_{N_k N_s} \: ,
\label{eq:fortyseven}
\end{equation}
\begin{equation}
W_{N_k N_s} = t^{N_k} \kappa^{N_s} \left[ \prod_{{\rm ( \: pairs \: of \:} \tau_s {\rm )} } 
G ( \tau_s - \tau_{s'} ) \right] \, e^{A_{\zeta}[{\bf r}(\tau),a(\tau)]} \: .
\label{eq:fortyeight}
\end{equation}
Compared to (\ref{eq:thirty}), the above expression involves additional multiple
integration over the times of orbital flips. Each term is associated with a spatial-orbital
path with $N_k$ kinks and $(N_s/2)$ $\eta$-phonon pairings, as shown in Fig.~\ref{fig:nine}(a). 
A Markov process is organized by inserting/removing spatial kinks and by attaching/removing 
the pairing lines. The acceptance rules for the pairing lines can be derived by extending 
the method of Section~\ref{sec:threeone} to two-time updates. For details see 
\cite{Kornilovitch2000}. The path shown in Fig.~\ref{fig:nine}(a) can be considered a 
space-time {\em diagram}. In fact, the update process just described is a version of the 
Diagrammatic Monte Carlo method \cite{ProkofevDMC}.

After the update rules are established, the JT polaron properties can be calculated with no 
approximations. The mass, spectrum, and density of states are obtained as for the conventional
lattice polarons. For the JT energy the thermodynamic estimator yields:
\begin{equation}
E_0 = \left\langle - \frac{N_k}{\beta} - \frac{\partial A_{\zeta}}{\partial \beta} 
- \frac{N_s}{\beta} - \frac{\hbar\omega}{\beta} \sum_{G}  
\frac{1}{G(\tau_s - \tau_{s'})} \frac{\partial G(\tau_s - \tau_{s'})}{\partial (\hbar\omega)}  
\right\rangle_{\rm shift}   .
\label{eq:fortynine}
\end{equation}
The first two terms are familiar from the Holstein case. The first is the kinetic energy of 
the electron. The second is the potential energy of interaction with the $\zeta$ plus the 
(positive) elastic energy of $\zeta$. The other two terms are new. The third one represents
the negative ``orbital energy'' caused by orbital flips. These processes excite phonons
$\eta$, which results in the positive fourth term. 

The number of excited phonons is an important characteristic because it provides an internal
consistency check for the algorithm: one expects {\em equal} mean number of $\zeta$ and
$\eta$ phonons. The phonon number estimator is derived as the derivative of the free energy 
with respect to $(\hbar\omega)$ under a fixed combination $\kappa^2/(M\omega)$. One obtains
\begin{equation}
N_{{\rm ph}, \zeta} = - \frac{1}{\beta} \left\langle \left. 
\frac{\partial A_{\zeta}}{\partial (\hbar\omega)} \right\vert_{\frac{\kappa^2}{M\omega}} 
\right\rangle_{\rm shift}   ,
\label{eq:fiftyone}
\end{equation}
\begin{equation}
N_{{\rm ph}, \eta} = - \frac{1}{\beta} \left\langle \sum_{G}  
\frac{1}{G(\tau_s - \tau_{s'})} \frac{\partial G(\tau_s - \tau_{s'})}{\partial (\hbar\omega)}  
\right\rangle_{\rm shift}   .
\label{eq:fiftytwo}
\end{equation}
\begin{figure}[b]
\centering
\includegraphics[width=11.7cm]{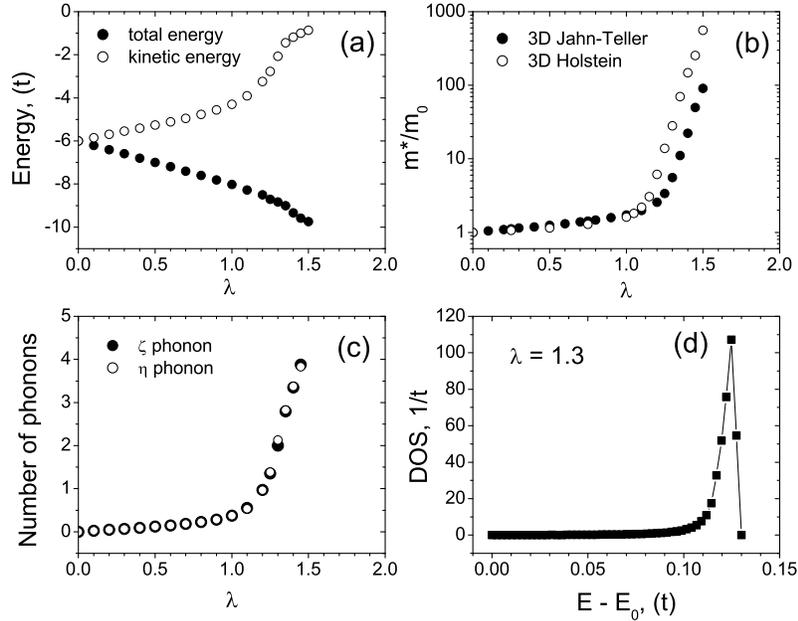}
\vspace{-0.5cm}
\caption{
Physical properties of the $d=3$ Jahn-Teller polaron at $\hbar\omega = 1.0 \, t$ 
\cite{Kornilovitch2000}.
(a) The total and kinetic energy. 
(b) The effective mass compared with the $d=3$ Holstein polaron at the same phonon frequency.
(c) The number of excited phonons of both types.
(d) The density of states of the JT polaron at $\lambda = 1.3$.  
}
\label{fig:JT_properties}       
\end{figure}

Results of PIQMC calculations are shown in Fig.~\ref{fig:JT_properties} \cite{Kornilovitch2000}.
Most properties behave similarly to those of the $d=3$ Holstein polaron at the same phonon
frequency, $\hbar\omega = 1.0 \, t$. For example, the kinetic energy, 
Fig.~\ref{fig:JT_properties}(a), sharply decreases by absolute value between $\lambda = 1.2$ 
and $1.4$. The JT polaron mass is slightly larger at the small to intermediate coupling, but 
several times smaller at the strong coupling. This non-monotonic behaviour of the ratio of
the JT and Holstein masses was later confirmed by accurate variational calculations   
\cite{ElShawish2003}, although in that work the JT polaron (and bipolaron) was investigated
in one spatial dimension. The relative lightness of the JT polaron is consistent with Takada's
result mentioned above \cite{Takada2000}. The number of excited phonons of both types is
shown in Fig.~\ref{fig:JT_properties}(c). As expected, numerical values coincide within the 
statistical error, which validates the numerical algorithm. Interestingly, the shape of the
phonon curves is similar to that of the logarithm of the effective mass. This suggests an 
intimate relationship between the two quantities, again similarly to the Holstein case.
Finally, the density of JT polaron states features the same peak at the top of the
band, caused by the spectrum flattening at large polaron momenta.  

In summary, the locality of the JT interaction and the independence of vibrating clusters  
result in the same extremity of polaron properties as in the $3d$ Holstein model. One should
expect that either a long-range JT or phonon dispersion will soften the sharp polaron features
and make them more ``normal''. This is an interesting research opportunity for the PIQMC 
method.

\section{ \label{sec:six}
Prospects
}

In this Chapter, a theoretical approach to lattice polarons based on the statistical path 
integral has been developed. A characteristic feature of the method is a tight integration of
analytical and numerical elements within one technique. The numerical part of the algorithm 
is greatly aided by the preliminary analytical steps: path integration over the ionic coordinates
and reduction of the double integral over imaginary time to a discrete sum over path segments.
Thus, on the way from the Hamiltonian to observables, exact analytical transformations are 
carried as far as possible increasing the accuracy and speed of the simulations. 

The PIQMC polaron method is still very much under development. Due to the space constraints,
only key elements have been presented here and many technical details have been left out.
This concluding section will be used to outline various extensions of the method. Some of them 
have already been done, others {\em can} be done but await proper implementation. Certain 
problems have not yet been solved even on the conceptual level, and those will be discussed too.

\subsection{ \label{sec:five}
Bipolaron 
}

The possibility of two polarons forming bound pairs, or bipolarons, opens a whole new 
dimension in polaron physics \cite{Alexandrov1995,Devreese1996}. The two major competing factors 
are the phonon-mediated attraction between the polarons that favours paring and direct Coulombic 
repulsion that prevents pairing. Kinetic energy, exchange effects and short-range lattice effects 
play important supporting roles that can tip the balance toward or against paring. All of this may 
lead to a rich and complex phase diagram. The bipolaron concept has been well-developed in the 
literature. The large continuous-space (Fr\"ohlich) bipolaron in ionic media was considered in
\cite{Verbist1990,Hiramoto1985,Emin1992}, and the small, localized, lattice bipolaron in 
\cite{Anderson1975}. Since the bipolarons are charged bosons at low density $n$, they can 
superconduct below the Bose-Einstein condensation temperature 
\begin{equation}
T_{\rm BE} = 3.31 \frac{\hbar^2}{k_B} \frac{n^{2/3}}{m^{\ast\ast}} \: .
\label{eq:fifty}
\end{equation}
The concept of {\em mobile} lattice bipolarons and bipolaronic superconductivity was put 
forward by Alexandrov and Ranninger in 1981 \cite{Alv81} and since then thoroughly 
developed by Alexandrov and co-workers 
\cite{Alexandrov1996,Kornilovitch2002,Alexandrov1992,AKM1996,Alv93,AlvM94,Alv03,Alexandrov2006}. 
Note that $T_{\rm BE}$ is inversely proportional to the bipolaron mass. This suggests the 
following recipe to increase the critical temperature according to the bipolaronic mechanism: 
{\em reduce the effective mass while keeping the bipolarons non-overlapping}. (Clearly, 
simply reducing the mass will increase the bipolaron radius and reduce the overlap density. 
There is a trade-off between the $n^{2/3}$ and $(m^{\ast\ast})^{-1}$ factors in 
(\ref{eq:fifty}), which creates a challenging optimization problem.)    
   
As with polarons, the first application of path integrals to the bipolaron was to the 
continuum-space Fr\"ohlich bipolaron \cite{Verbist1990,Hiramoto1985}. Until very recently, 
the only application of path integrals to the lattice bipolaron was the work by de Raedt and 
Lagendijk \cite{DeRaedtBipolaron} on the discrete-time version of their path-integral QMC 
method. The authors limited the consideration to the extreme adiabatic limit and computed 
the phase diagram of the Holstein bipolaron with an additional Hubbard repulsion. 
A continuous-time version of PIQMC was developed very recently and the first results 
presented in \cite{Hague2006}.    

The lattice bipolaron, mostly with Holstein interaction, has been studied by a variety of
other numerical methods: 
exact diagonalization \cite{Wellein1996,Weisse2000}, 
variational \cite{Bonca2000,Bonca2001,ElShawish2003},
density matrix renormalization group \cite{Zhang1999}, 
diagrammatic quantum Monte Carlo \cite{Macridin2004},
and Lang-Firsov quantum Monte Carlo \cite{Hohenadler2005,Hohenadler2005b}.  

Conceptually, generalizing the PIQMC method to the bipolaron is straightforward. There are 
two paths that visualize the imaginary-time evolution of the two fermions. To enable calculation 
of the effective mass, open boundary conditions in imaginary time must be used: the top 
ends of the paths must be the same as the bottom ones up to an arbitrary lattice translation
$\Delta {\bf r}$. Singlet and triplet states can be separated by allowing the paths to exchange,
see Fig.~\ref{fig:nine}(b). Then the bipolaron spectrum, mass, and singlet-triplet splitting is 
calculated as explained in Section~\ref{sec:two}. Phonon integration is performed as before with 
the same resulting action, (\ref{eq:sixteen})-(\ref{eq:twenty}). The difference is the force 
that is acting on the oscillator ${\bf m}$ at time $\tau$. In a bipolaron system it receives 
contribution from both electrons, 
$f_{\bf m}(\tau) = f_{\bf m}[{\bf r}_1(\tau)] + f_{\bf m}[{\bf r}_2(\tau)]$. Since the action 
is quadratic in $f$, it becomes a sum of three terms that contain the products 
$f[{\bf r}_1(\tau)] f[{\bf r}_1(\tau')]$, $f[{\bf r}_2(\tau)] f[{\bf r}_2(\tau')]$, and 
$f[{\bf r}_1(\tau)] f[{\bf r}_2(\tau')]$, respectively. The first two terms describe the 
retarded self-interaction of the two paths and are responsible for polaron formation. The third 
cross-term describes the interaction between the paths. It is responsible for the interaction 
between the polarons and for bipolaron formation. This is illustrated in Fig.~\ref{fig:nine}(b). 
The estimators for various observables and specific expressions for the segment-to-segment 
contributions are derived analogously to the polaron case. Technical details of the derivation 
have not yet been published, and will be published elsewhere.

One interesting effect that has already been considered by the PIQMC method is the ``crab-like'' 
bipolaron that can exist on certain lattices \cite{Hague2006}. Usually, the bipolaron is much 
heavier that the constituent polarons, its mass scaling as the second power of the polaron mass 
in the non-adiabatic regime and as the fourth power in the adiabatic limit \cite{Alexandrov1986}. 
(Path integrals provide a useful visualization of this effect, see Fig.~\ref{fig:nine}(b). At a 
small phonon frequency, the two paths interact over their entire length. Therefore they are much 
more difficult to separate, which results in slower imaginary-time diffusion, and hence a heavier 
particle.) However, on the triangular, face-centered cubic and some other lattices the 
{\em intersite} bipolaron can move without breaking. This results in an effective mass that 
scales only {\em linearly} with the polaron mass. This effect was predicted some time ago by 
Alexandrov \cite{Alexandrov1996}, and recently confirmed by exact PIQMC simulations in
\cite{Hague2006}.

\subsection{ \label{sec:fivetwo}
Further extentions
}

Recall that the basic Hamiltonian (\ref{eq:fifteen}) contains the simplest possible form of the
electron kinetic energy and the simplest possible model of the crystal lattice. It will now be 
explained how to relax these restrictions. First of all, it is straightforward to add electron 
hopping beyond the nearest neighbours. Because the path is defined in real space, this 
only requires the introduction of additional kink types that move the path in the respective 
directions. The values of the distant hopping integrals can be absolutely independent of the
nearest-neighbour values as long as they are negative. The methods of Section~\ref{sec:threeone} 
is readily generalized to paths containing kinks of different magnitude. The statistical weight 
(\ref{eq:twentythree}) is now a product of factors $(t_i \Delta \tau)$ along the path. The stochastic 
acceptance rules now contain the factor $(t_i \beta)$ where $t_i$ is the hopping amplitude of the 
proposed kink. It is interesting though that contribution to the kinetic energy is the {\em same} 
(equal to $\beta^{-1}$) for all kinks independently of the $t_i$ value. It is the mean number of 
kinks that is affected by $t_i$, leading to different partial kinetic energies along different 
lattice directions. In exactly the same manner one can study models with anisotropic 
(but nearest-neighbour) electron hopping. This was done in \cite{Kornilovitch1999,anisotropy},
cf. Section~\ref{sec:fourtwo}. There remains one important restriction on the values of the 
hopping amplitudes: they all must be negative to ensure the positivity of the path's statistical 
weight. If even a single $t_i$ is positive, some paths will continue being positive but {\em some} 
will have negative weights. That will introduce a sign-problem, which will eventually render the 
simulation numerically unstable. Another extension related to the kinetic energy is lattices of 
different symmetries. Again, all that is changed is the table of kink types that specifies which 
two spatial points (lattice sites) each particular kinks connects. In this way, PIQMC has been 
applied to the triangular, face-centred-cubic, and hexagonal Bravais lattices in \cite{Hague2005}.

It should be mentioned that PIQMC can also simulate the (bi)polaron in dimensionalities larger
than 3, see Fig.~\ref{fig:Holstein}. (The $d=4$ Holstein polaron was previously investigated in 
\cite{Ku2002}.) Although such calculations have little practical value, they can still be useful 
in assessing the accuracy of approximate methods that rely on the number of spatial dimensions 
being large. (An example of such a method would be the Dynamical Mean Field Theory 
\cite{Capone2003}.) 
  
Consider now the phonon subsystem. It is well-known that a bosonic path integral can be calculated
analytically for any quadratic action. This means that the ionic coordinates can be eliminated 
even when they are coupled, that is in the case of dispersive phonons. Moreover, integration can
be done for any number of degrees of freedom per unit cell. De Raedt and Lagendijk studied a
one-dimensional Holstein polaron with phonon dispersion as early as in 1984 \cite{DeRaedt1984}.
They found that the critical coupling of polaron formation decreases as the phonon mode becomes
soft. To our knowledge, this remains the only exact analysis of a model with dispersive phonons
to-date. A general phonon integration for the purposes of PIQMC method was performed in 
\cite{multiphonon}. Conceptually, the result is similar to (\ref{eq:sixteen})-(\ref{eq:thirty}). 
The action is a sum of a periodic contribution and a correction due to the shifted boundary 
conditions. The action is a double integral over the imaginary time. However, there is an 
additional sum over the phonon momentum and branches. In addition, the action involves the 
eigenvalues and eigenvectors of the dynamical matrix. Thus, the polaron action comprises full 
information about the phonon spectrum and polarizations.

An important advantage of the PIQMC method is the ability to study temperature effects. At a 
finite temperature $T$, the projected partition function $Z_{\bf K}$ receives contributions
from high-energy states with momentum ${\bf K}$. It is therefore not meaningful to
calculate the polaron mass and spectrum. Instead, the conventional periodic boundary conditions 
in imaginary time must be used, which makes all states contribute to the full thermodynamic
partition function $Z$. The parameter $\Delta {\bf r}$ is set to zero. The remaining periodic 
action (\ref{eq:seventeen}) is valid at any temperature, large or small. This enables 
approximation free calculation of thermodynamic properties of the polaron: the internal 
energy, specific heat, number of excited phonons, and static e-ph correlation functions. 

The grand challenge for PIQMC, as for almost any Quantum Monte Carlo method is efficient simulation
of many-body systems. If the number of particles is three or more, the {\em fermionic} sign-problem
is in general unavoidable. It is possible that the sign-problem in many-fermion e-ph models will be 
less severe that in the case of repulsive electron-electron models. Consider the limit of low 
electron density and strong e-ph interaction. The polarons will be bound in bipolarons, and 
the paths will fluctuate in pairs. To account for statistics, the top path ends must 
be constantly permuted, with even/odd permutations contributing $+1/-1$ phase factor to the partition 
function. However, any odd permutation will have to separate a path pair. This will increase the 
energy of the configuration and such an update will likely be rejected. In contrast, {\em some} 
even permutations will exchange whole bipolarons, which will keep the paths together with no 
significant energy increase. Such updates will likely be accepted. As a result, there will be many 
more even than odd permutations, and the average sign will be well defined. The stronger the 
coupling the more pronounced this effect will be, and the more the system will approach a purely 
bosonic limit (which is sign-problem-free).

The temperature-control capability offers another unique opportunity: calculation of the Meissner
fraction and superconducting critical temperature. The method was devised by Scalapino et al
\cite{Scalapino1993} and is based on the sum rule for the off-diagonal current-current 
correlation function. By comparing its static limit with the kinetic energy, one could determine 
a temperature at which the two quantities start to differ. This signifies the appearance of a 
Meissner fraction and determines the $T_c$.

Finally, we comment on the calculation of dynamical (bi)polaron properties such as the spectral
function or optical conductivity. This is a classic and difficult problem for most existing
Monte Carlo methods. Derivation of real-time correlators from imaginary-time correlators amounts
to solving an ill-posed inversion problem, which is usually regularized by methods such as
maximum entropy \cite{Jarrell_1996}, Pade approximants, or stochastic optimization 
\cite{Mishchenko2000}. Any of this techniques can be used in conjunction with the PIQMC method.

\subsection{ \label{sec:fivethree}
Conclusions
}

In summary, path integrals have played and continue to play an important role in the development
of polaron physics. The path-integral quantum Monte Carlo method has proven to be a powerful 
and versatile tool. It has some unique advantages, such as system-size independence, the ability
to calculate the density of states, mass isotope exponents, and temperature effects. Compared to
the exact diagonalization, density-matrix renormalization group, or variational methods, PIQMC 
has a larger statistical error, but provides unbiased estimates for the (bi)polaron 
properties. Several novel qualitative results have been obtained with the PIQMC method: small 
polaron mass in models with long-range electron-phonon interaction, enhancement of the 
anisotropy of the polaron spectrum, a peak in the polaron density of states in the adiabatic 
limit, and the isotope effect on the polaron spectrum. These and some other effects have been 
discussed in this Chapter in detail.

\section*{Acknowledgements}

It is my pleasure to thank Sasha Alexandrov for long-standing collaboration and motivation
for this research. I also thank James Hague and John Samson for useful discussions on the subject 
of this Chapter, and Natalia Kornilovich for help with organizing numerical data. I am grateful
to James Hague for help with figure~\ref{fig:IsotopeMass}. This work has been supported by 
EPSRC (UK), grant EP/C518365/1.

%
%
%
%
%
%
%
%
%
%

%
%



\printindex
\end{document}